%% file: main.tex
\documentclass[manuscript]{acmart}

\usepackage{graphicx}   % 基础图形支持
\usepackage{subcaption} % 子图支持
\usepackage{float}      % 精确控制图片位置[htbp]
\usepackage{xcolor}
\usepackage{soul}
\acmConference[CHI '26]{CHI conference on Human Factors in Computing Systems}{April 13--17, 2026}{Barcelona, Barcelona}

\begin{document}

\title{The Wireless Charger as a Gesture Sensor: A Novel Approach to Ubiquitous Interaction}

\input{authors}

\input{abstract}


\maketitle

\input{introduction}

\input{relatedwork}

\input{preliminary}

\input{framework}

\input{method}

\input{prototype}

\input{evaluation}

\input{userstudy}

\input{discussion}

\input{conclusion}

% \printbibliography
\bibliographystyle{ACM-Reference-Format}
\bibliography{references}

\end{document}

%% file: authors.tex
% ================== 作者信息开始 ==================
% 注意：在 anonymous 选项下，这些作者信息会自动被隐藏，不会出现在匿名稿里。
% 录用后只需把 \documentclass 里的 anonymous 删掉，作者就会正常显示。

\author{Weiyi Wang}
\email{weiyi_wang@sjtu.edu.cn}
\affiliation{%
  \institution{Shanghai Jiao Tong University}
  \city{Shanghai}
  \country{China}
}

\author{Lanqing Yang}
\email{yanglanqing@sjtu.edu.cn}
\affiliation{%
  \institution{Shanghai Jiao Tong University}
  \city{Shanghai}
  \country{China}
}

\author{Linqian Gan}
\email{leo-gan@sjtu.edu.cn}
\affiliation{%
  \institution{Shanghai Jiao Tong University}
  \city{Shanghai}
  \country{China}
}

\author{Guangtao Xue}
\email{gt_xue@sjtu.edu.cn}
\affiliation{%
  \institution{Shanghai Jiao Tong University}
  \city{Shanghai}
  \country{China}
}

% 可选：短作者列表（出现在页眉）
% \renewcommand{\shortauthors}{First Author et al.}
% ================== 作者信息结束 ==================

%% file: abstract.tex
\begin{abstract}
\textbf{Abstract} Advancements in information technology have increased demand for natural human–computer interaction in areas such as gaming, smart homes, and vehicles. 
However, conventional approaches like physical buttons or cameras are often limited by contact requirements, privacy concerns, and high costs.
Motivated by the observation that these EM signals are not only strong and measurable but also rich in gesture-related information, we propose \textit{EMGesture}, a novel contactless interaction technique that leverages the electromagnetic (EM) signals from Qi wireless chargers for gesture recognition. 
\textit{EMGesture} analyzes the distinctive EM features and employs a robust classification model. 
The end-to-end framework enables it capable of accurately interpreting user intent. 
Experiments involving $30$ participants, $10$ mobile devices, and $5$ chargers showed that \textit{EMGesture} achieves over $97$\% recognition accuracy. 
Corresponding user studies also confirmed higher usability and convenience, which demonstrating that \textit{EMGesture} is a practical, privacy-conscious, and cost-effective solution for pervasive interaction.

\end{abstract}

%% file: introduction.tex
\section{Introduction}
With the rapid advancement of information technology, the demand for human–machine interaction products has been steadily increasing, driving continuous growth in the global HMI market. According to reports from People’s Daily, consumers increasingly prefer products that feature natural and intelligent interaction, making the convenience and intelligence of interaction an important consideration in purchasing decisions ~\cite{Smarthom87:online}. Market forecasts further validate this trend. According to Statista, the global human–machine interface market is projected to reach USD 7.24 billion by 2026, nearly double its size in 2020, with a compound annual growth rate (CAGR) exceeding 10\% ~\cite{Globalhu52:online}. Meanwhile, the broader human–machine interaction industry is expected to grow to USD 934.84 billion, with a CAGR of 18.5\% ~\cite{HumanCom90:online}.

Despite this momentum, current interaction modalities remain limited. Broadly, they can be categorized into contact-based and contactless approaches. Contact-based interaction is the mainstream form, including: (1) Touchscreen interaction, known for its accuracy and user familiarity, but limited in humid or oily environments ~\cite{Whatares45:online}, with relatively high costs and technical complexity ~\cite{65Intera48:online}. (2) Keyboard and mouse interaction, which provides efficiency but requires external devices. (3) Button-based interaction, which is simple but restricted in functionality ~\cite{ThePower76:online}. In contrast, contactless methods include: (1) Voice interaction, which supports a limited number of devices and struggles with complex command comprehension, while also raising concerns about voice-privacy leakage ~\cite{TheVoice1:online}. (2) Wi-Fi-signal-based interaction, which remains hindered by low accuracy ~\cite{HowtoChe84:online}. (3) Camera-based interaction, which consumes high power ~\cite{sanmiguel2016energy}, lacks universality, and poses privacy risks due to its reliance on image data access ~\cite{IoTSecur47:online}.

These limitations become particularly evident in diverse application scenarios. For instance, in driving environments, contact-based operations may distract drivers and increase risks. According to the U.S. National Highway Traffic Safety Administration, distracted driving led to 3,275 fatalities in 2023, averaging nearly nine deaths per day ~\cite{Distract53:online}. In kitchen environments, touchscreen devices often fail due to water or grease on the hands ~\cite{1Whydoes87:online}. In smart home scenarios such as remote appliance control, users still rely heavily on physical remotes, whereas more intelligent interaction modes are anticipated ~\cite{PDFSmart27:online}. Likewise, contactless methods remain insufficient: most voice interaction systems only support simple one-turn commands and fail to align with long-term user habits.

To address these challenges, this paper proposes a novel contactless gesture interaction method that leverages the electromagnetic (EM) signals emitted during wireless charging. Our experiments reveal that, based on the negative feedback mechanism of wireless chargers ~\cite{yang2024privacy}, such chargers emit strong, frequency-modulatable, measurable, and stable EM signals that can be captured by antennas (details are provided in Section ~\ref{sec:feedback}). When a user performs a gesture, the collected EM signals exhibit distinctive variations. Building on this observation, we introduce \textit{EMGesture}, a system that exploits wireless charging EM signals to achieve gesture recognition. Specifically, we decouple the EM radiation characteristics of Qi-standard chargers and establish a dynamic mapping model between gesture-induced disturbances and EM responses. \textit{EMGesture} uses antennas to collect signals, which are amplified by universal signal amplifiers and then input into the HackRF software-defined radio platform for data collection. Environmental EM noise is mitigated using spectral subtraction combined with Variational Mode Decomposition (VMD), resulting in clean training samples. Finally, a Random Forest classifier is trained, achieving 97.59\% accuracy in recognizing nine basic gestures across 30 participants, 10 device models, and 5 charger models. Compared to existing interaction methods, our approach offers significant advantages, including reduced spatial constraints, broad applicability, enhanced privacy protection, low cost, low power consumption, and alignment with user habits. Furthermore, extensive comparative and ablation studies confirm its robustness across diverse users, devices, and environments.

% Looking ahead, with the continuous advancement of built-in magnetometers and antennas in smartphones and wireless chargers, the proposed gesture interaction method can be further simplified, eventually relying directly on built-in sensors for implementation. This evolution will greatly enhance its practicality and application potential in everyday scenarios.

The main contributions of this work are summarized as follows:
\begin{itemize}
\item We propose \textit{EMGesture} for the first time, which is a novel contactless interaction technique that leverages the electromagnetic (EM) signals inherently emitted from Qi-standard wireless chargers for gesture recognition, enabling mid-air interaction without requiring any hardware modifications to smartphones.
\item We develop a comprehensive gesture feature extraction and processing pipeline, which includes receiving and amplifying gesture signals, employing a short-window FFT to generate an average power spectrum over time as a stable gesture feature, and effectively reducing environmental EM noise using Variational Mode Decomposition (VMD).
\item We design a robust gesture classifier capable of accurately distinguishing between eight fundamental gestures and the idle (charging-only) state, achieving a high recognition accuracy of $97.59$\% for nine categories.
\item We demonstrate the practicality and effectiveness of our system through large-scale experiments involving 30 participants, 10 different device models, and 5 charger models, achieving an overall recognition accuracy of over $88$\%. Corresponding user studies further confirmed the system's high usability and convenience.

\end{itemize}

%% file: relatedwork.tex
\section{Related Work}
\textbf{Electromagnetic Signals Based Intelligent Sensing.} Recently, many schemes for using electromagnetic signals for interaction have been proposed. Magstroke ~\cite{abdelnasser2020magstroke} uses a mobile device's magnetometer to sense the motion trajectory of a magnet worn on the user's hand, proposing a virtual keyboard based on magnetic field perception. MAGNECOMM+ ~\cite{xue2021magnecomm+} also uses a device's magnetic sensor, but to receive EM signals generated by a computing device to enable data communication. CamRadar ~\cite{liu2023camradar} detects the presence of cameras by sensing electromagnetic radiation leaked from their components, proposing a novel hidden camera detection system. Zhang et al. ~\cite{zhang2024adaptive} achieve unique device identification by analyzing hardware-level fingerprints within their radiated EM emissions.
Compared with prior work that directly exploits EM signals emitted by specific devices, we observe that wireless chargers are far more ubiquitous in everyday environments and naturally emit stronger EM fields. However, no existing research has investigated using the EM signals radiated from wireless chargers for gesture interaction. Moreover, our observations reveal that due to the negative feedback mechanism inherent in wireless charging, it is possible to artificially generate controllable and stable EM signals (details are provided in Section ~\ref{sec:feedback}), which greatly benefits feature extraction. 

\textbf{Gesture Recognition.} 
A significant body of prior work has investigated gesture sensing through on-body and off-body modalities. 
For example, GestureTrack~\cite{khanna2024hand} enabled gesture recognition for blind users by relying solely on smartwatch gyroscopes. 
Other systems have leveraged acoustic signals: MAF~\cite{yang2024maf} employed surface and leakage acoustic waves from bone-conduction headphones, while SpeakerGesture~\cite{li2022room} used chirp-based acoustic sensing to achieve room-scale gesture recognition. 
Specialized radio-frequency (RF) hardware has also been explored. GesturePrint~\cite{xu2024gestureprint} combined millimeter waves (mmWave) with user identity binding, while RadarNet~\cite{hayashi2021radarnet} and Rodar~\cite{jin2024rodar} achieved gesture recognition using micro-radar and mmWave sensors. 
To reduce hardware dependency, commodity Wi-Fi has been widely adopted. WiGesture~\cite{gao2021towards} introduced cross-location gesture recognition with position-independent features, and subsequent systems such as WiGesID~\cite{zhang2022wi} and UniFi~\cite{liu2024unifi} exploited Wi-Fi Channel State Information (CSI) to jointly model gestures and user identity, while also adapting sensing across new environments.  
Despite these advances, each approach exhibits trade-offs. Wearable and some acoustic systems are not truly contactless, requiring dedicated devices such as watches or headphones. Specialized RF systems deliver high accuracy but depend on costly and non-ubiquitous hardware. Commodity Wi-Fi solutions improve deployment feasibility but often suffer from instability and limited resolution. 
Our work addresses these gaps: unlike wearable approaches, our method is completely device-free; unlike radar-based solutions, it repurposes an already ubiquitous, low-cost device available in homes, offices, and vehicles—the wireless charger. 
Furthermore, our classifier (Random Forest) is computationally lightweight, avoiding the overhead of deep learning, while inherently preserving privacy by capturing no audio or visual data.

\textbf{Wireless Chargers Related Research.} The Qi standard, introduced by Wageningen et al.~\cite{van2010qi}, established the fundamental communication protocols for wireless charging. 
Subsequent research has extended wireless charging into new domains such as automotive applications; for example, Mohamed et al.~\cite{mohamed2024wireless} proposed in-vehicle wireless charging solutions. It also provides ideas for the final use scenario of \textit{EMGesture} to expand from mobile device wireless chargers to other wireless chargers.
Beyond power transfer, researchers have begun to harvest information from the EM fields generated during charging. 
Zhan et al.~\cite{zhan2024voltschemer} demonstrated a security attack by injecting inaudible voice commands through magnetic field interference. 
MagID~\cite{li2024magnetic} employed a magnetic sensor array to authenticate device identity during charging, while Ni et al.~\cite{ni2023uncovering} leveraged coil ringing and field disturbances to infer user interaction with smartphones.  
These studies confirm that wireless charging fields contain rich, harvestable data. 
However, existing work has focused primarily on adversarial security applications~\cite{zhan2024voltschemer}, device authentication~\cite{li2024magnetic}, or device-centric sensing~\cite{ni2023uncovering}. 
To our knowledge, no prior work has systematically decoupled the ambient EM field perturbations from standard Qi chargers to model and classify a vocabulary of contactless, mid-air gestures. 
This paper demonstrates that the wireless charger itself can be a robust and practical gesture sensor.

%% file: preliminary.tex
\section{Preliminary Study}
Our preliminary experiments are designed to answer two fundamental questions: 
i) Does a wireless charger generate strong electromagnetic (EM) signals while charging a device? 
ii) Do different user gestures induce measurable perturbations in these EM signals? 
The answers to these questions are crucial for validating the feasibility of gesture interaction based on wireless charging EM emissions.

\begin{figure}[tbhp]
    \centering
    \begin{subfigure}{0.23\textwidth}
      \centering
      \includegraphics[width=\linewidth]{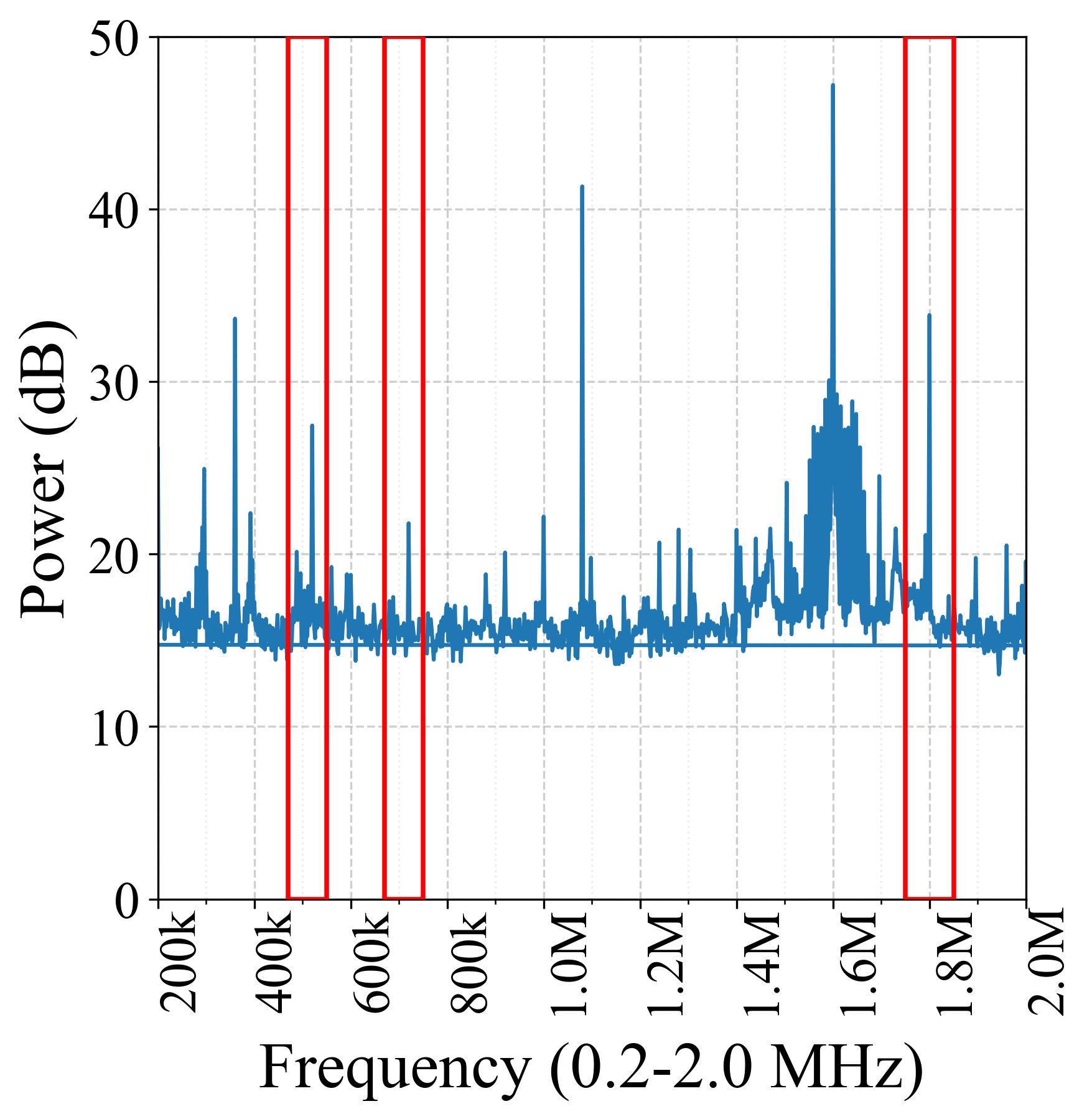}
      \caption{Charging}
      \label{Charging}
    \end{subfigure}
    \hfill
    \begin{subfigure}{0.23\textwidth}
      \centering
      \includegraphics[width=\linewidth]{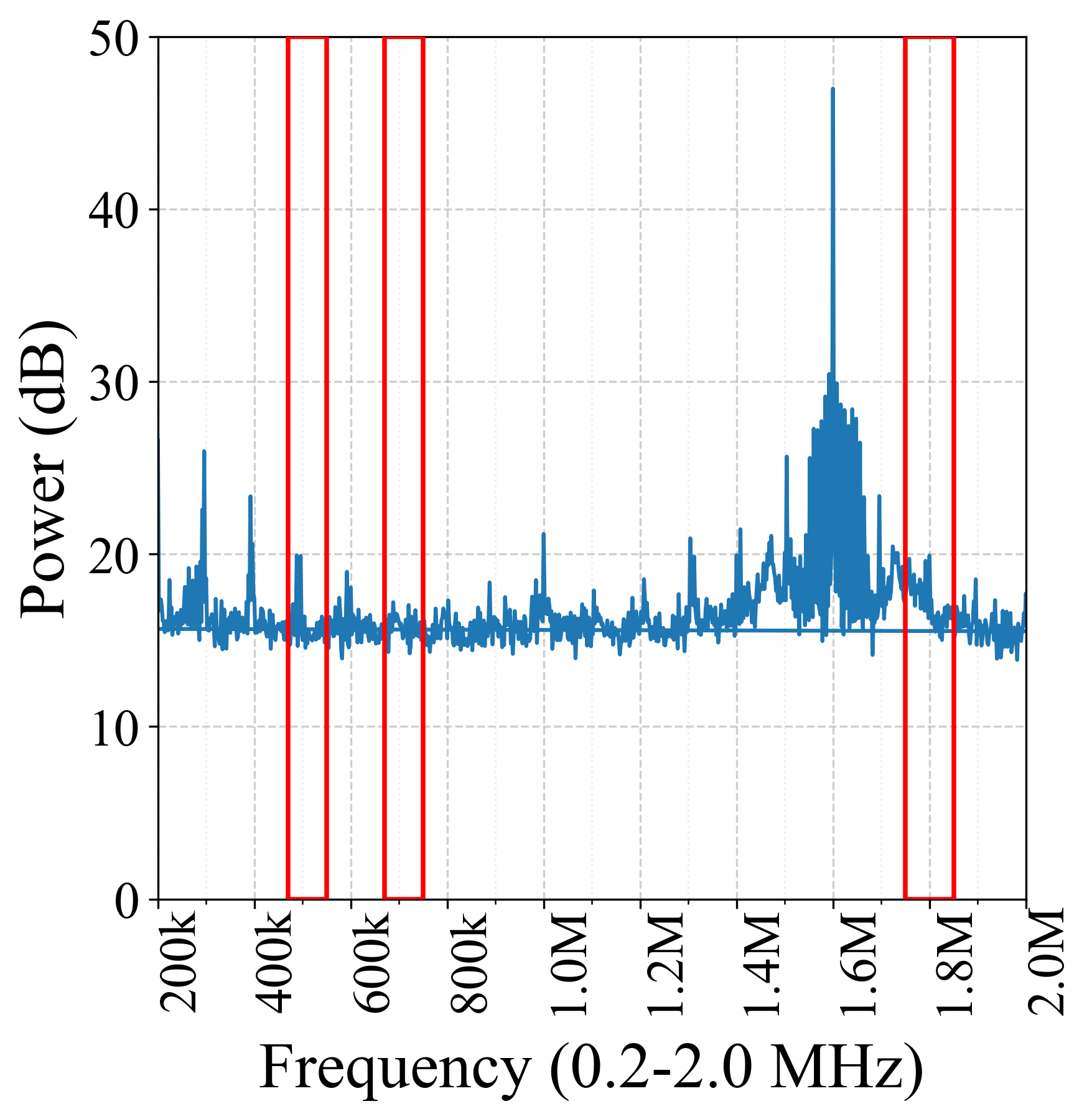}
      \caption{Without charging}
      \label{WithoutCharging}
    \end{subfigure}
    \hfill
    \begin{subfigure}{0.23\textwidth}
      \centering
      \includegraphics[width=\linewidth]{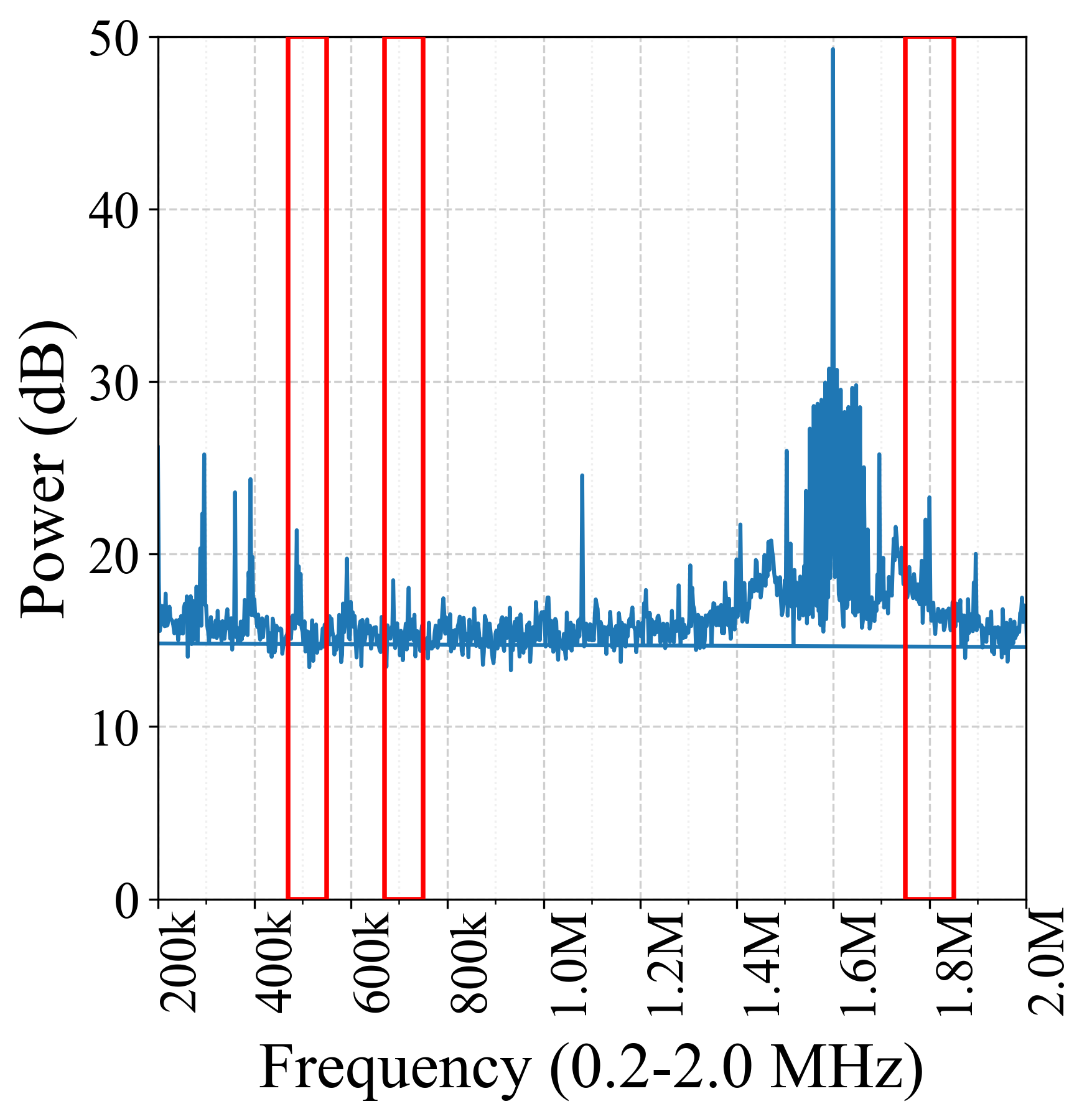}
      \caption{Hand-spreading}
      \label{HandSpreading}
    \end{subfigure}
    \hfill
    \begin{subfigure}{0.23\textwidth}
      \centering
      \includegraphics[width=\linewidth]{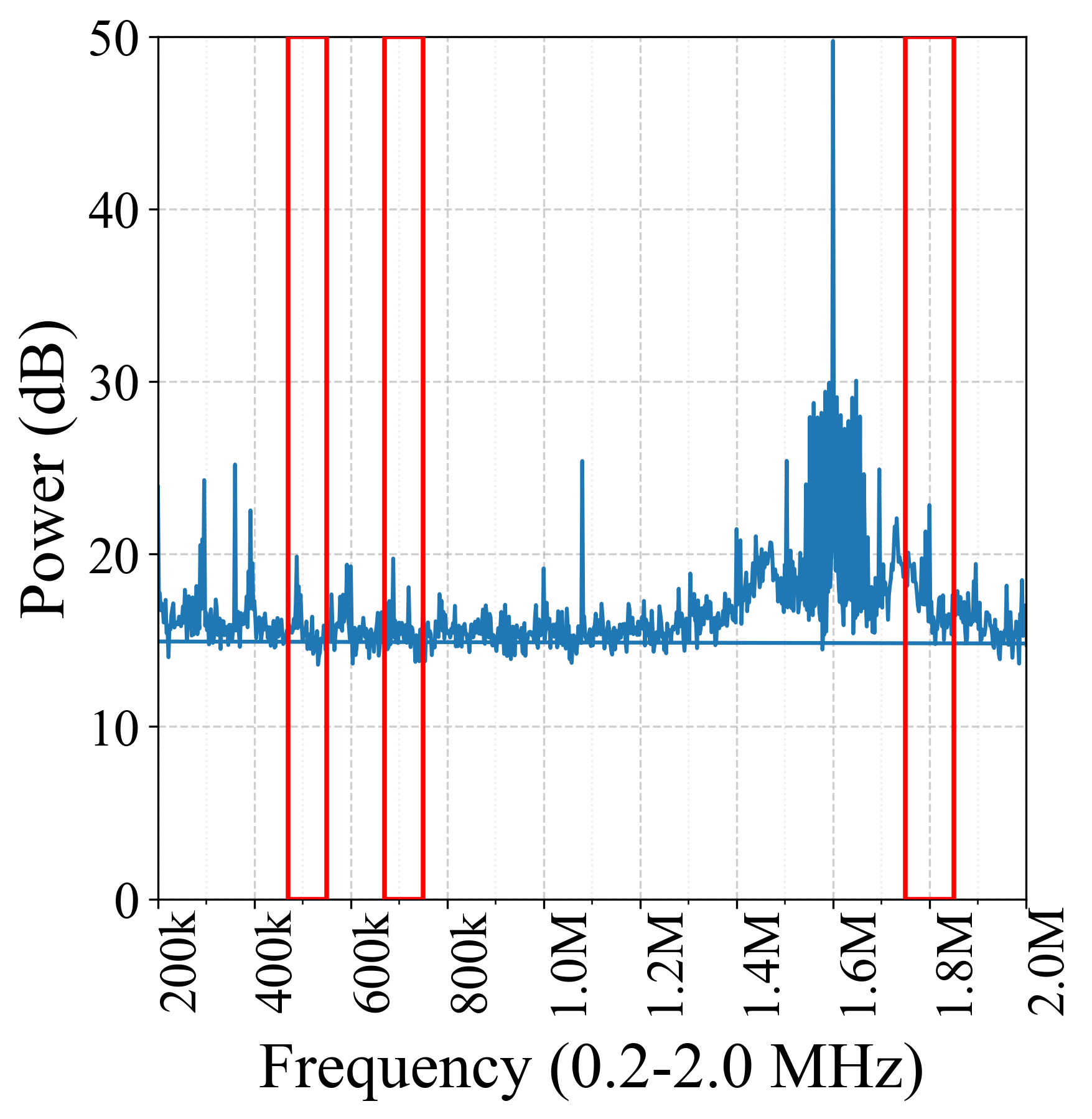}
      \caption{Number 1}
      \label{Number1}
    \end{subfigure}
    \caption{Electromagnetic signals under different wireless charger states and different gestures}
    \label{different_signal_and_gesture}
\end{figure}

\textbf{i) Does a wireless charger generate strong EM signals during charging?}  
We collected EM signals from a smartphone under two conditions: (1) actively charging via a wireless charger, and (2) idle (not charging). 
The corresponding frequency-domain representations are shown in Fig.~\ref{Charging} and Fig.~\ref{WithoutCharging}. 
By comparing the two cases, we observe that strong and stable EM signals appear across multiple narrow frequency bands only when the device is actively charging. The frequency band with obvious signal is framed in red.
These signal frequency bands have obvious gesture characteristics, making them promising candidates for feature extraction in gesture recognition tasks.

\textbf{ii) Do different user gestures affect EM signal characteristics?}  
To investigate this, we recruited a volunteer to perform several distinct hand gestures over an active wireless charger, while recording the resulting EM signals.  
Representative results are shown in Fig.~\ref{HandSpreading} and Fig.~\ref{Number1} for the \textit{hand-spreading} and \textit{number 1} gestures, respectively. 
We observe that each gesture introduces unique perturbations into the EM signals. 
In particular, the intensity of signals across multiple frequency bands attenuates to varying degrees depending on the specific gesture. 
These distinct, gesture-dependent perturbations confirm the potential to classify and recognize different gestures based on wireless charging EM signals.

%% file: framework.tex
\section{Framework}
\begin{figure}[tbp]
  \centering
  \includegraphics[width=0.8\linewidth]{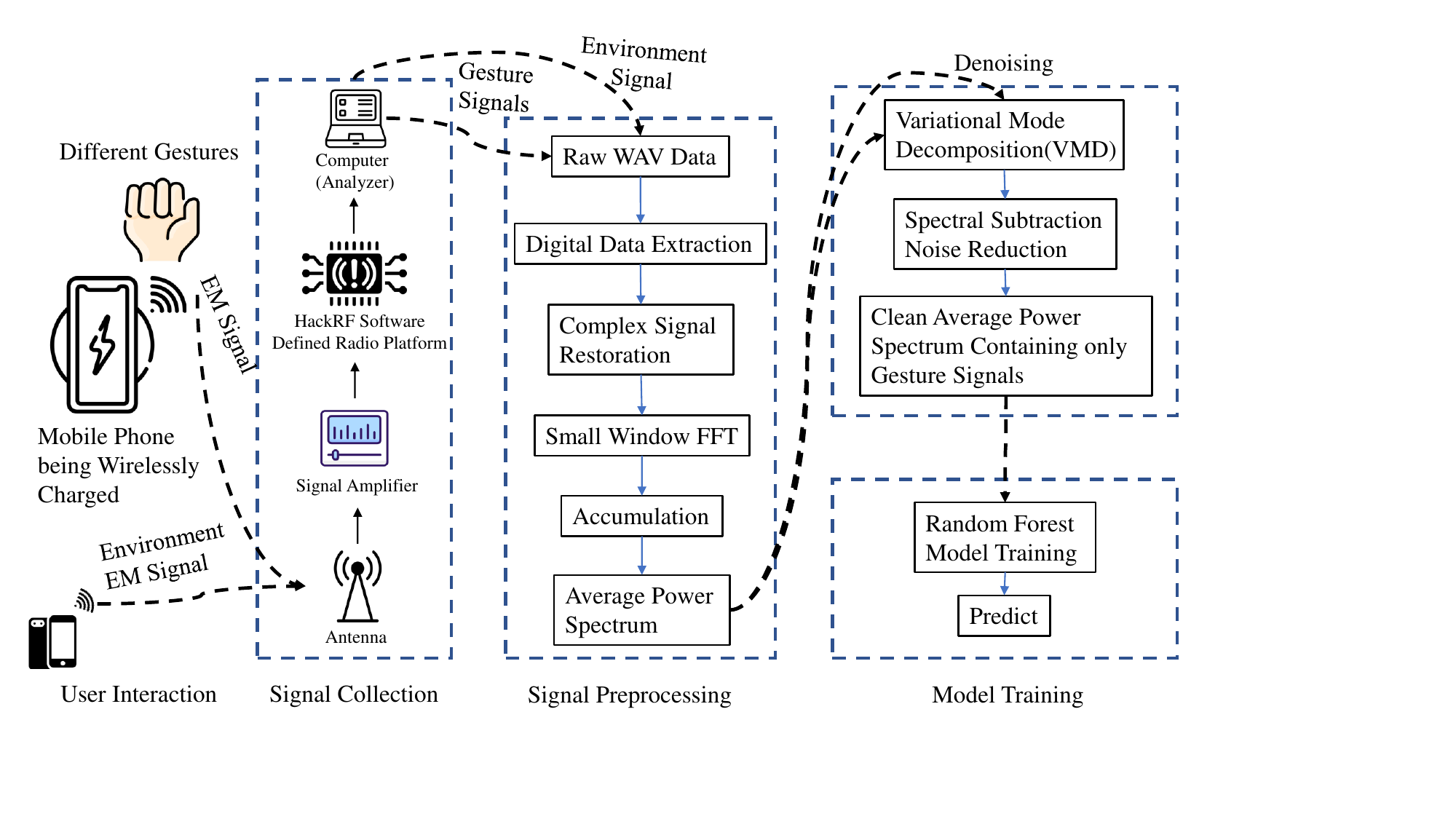} 
  \caption{\textit{EMGesture} overall framework diagram}
  \label{system}
\end{figure}
The overall architecture of our system is shown in Fig.~\ref{system}. It consists of five primary components: user interaction, signal collection, signal preprocessing, denoising, and model training.

\textbf{User Interaction \& Signal Collection.} 
In the user interaction module, a wireless charger with a mobile phone being charged emits strong electromagnetic signals (including signals during Qi protocol communication, charging field, and power supply signals), which are perturbed by the user's hand gestures. In the signal collection module, an antenna captures these nearby electromagnetic field signals (containing both the gesture-perturbed signal and ambient noise). The antenna converts the EM field into a voltage signal, which is then amplified. The signal is processed by the HackRF software-defined radio (SDR) platform ~\cite{palama20245g}, which saves the raw signal data for the preprocessing module.

\textbf{Signal Preprocessing \& Denoising.} 
In the preprocessing module, the \textit{i} and \textit{j} components of the raw signal are extracted to reconstruct the complex signal. Each segment (corresponding to one gesture sample) is then divided into multiple non-overlapping short windows.  
An FFT is applied to each window, and the resulting spectra are accumulated and averaged to produce a stable Average Power Spectrum for that segment.  
This process is carried out for both gesture-perturbed signals and independently recorded ambient noise.  
To reduce noise interference, we apply spectral subtraction: the averaged noise spectrum is subtracted from the averaged gesture spectrum, yielding a clean spectral representation that retains only gesture-relevant features.

\textbf{Model Training \& Classification.}
Finally, this clean spectral data is divided into training and validation sets. These feature sets are fed into a random forest model to be trained and to classify different gestures, completing the system design. The specific details of each module will be discussed in the following sections.

%% file: method.tex
\section{Method}
\subsection{User Interaction}
Our experimental setup consists of a standard Qi wireless charger placed on a tabletop, with a smartphone actively charging on it. 
A sensing antenna is positioned approximately 25 cm directly above the phone. 
Users perform hand gestures in the mid-air space between the charger and the antenna. 
These gestures perturb the otherwise stable electromagnetic (EM) field generated by the wireless charger. 
Since different gestures introduce distinct disturbances, they result in measurable and characteristic variations in the EM signals captured by the antenna ~\cite{ma2023mimo}.

\subsection{Signal Collection}
The signal collection pipeline begins with the antenna, which converts the perturbed EM field into an analog voltage signal. 
This signal is immediately passed through a general-purpose amplifier to boost its strength. 
The amplified signal is then fed into the HackRF software-defined radio (SDR) platform, which is connected via USB to a computer running SDRSharp software. 
The SDRSharp software will sample and digitize the signal and store the raw data as \texttt{.wav} files for signal preprocessing stage. It is worth noting that at this stage, in addition to collecting the electromagnetic signals of each gesture, environmental noise signals must also be collected for denoising stage.

\subsection{Signal Preprocessing}
\begin{figure}[tbp]
  \centering
  \includegraphics[width=0.7\linewidth]{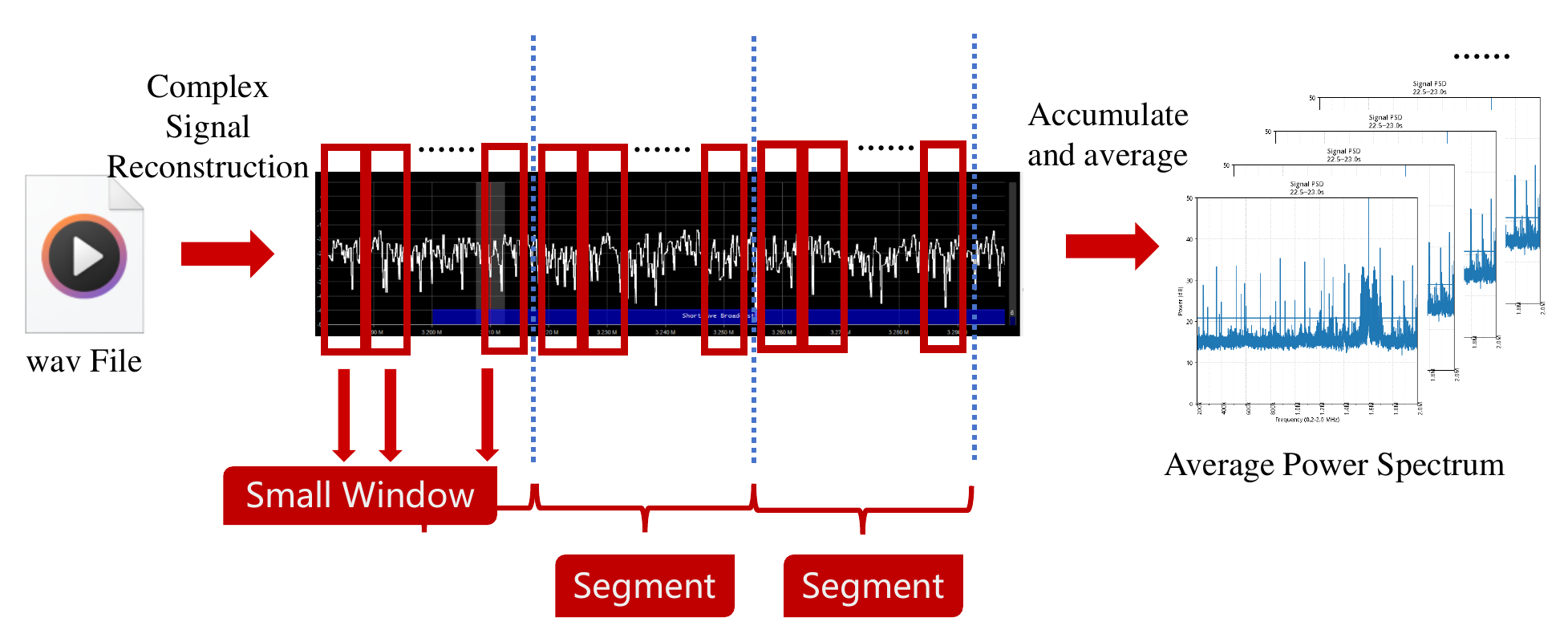} 
  \caption{Average power spectrum solution process}
  \label{fig:AverageSpectrum}
\end{figure}

Signal preprocessing consists of two steps: complex signal reconstruction and feature extraction. The flowchart of this stage is shown in Fig.~\ref{fig:AverageSpectrum}

\textbf{Complex Signal Reconstruction.} 
We first read the \texttt{.wav} files to obtain the in-phase \textit{i} and quadrature \textit{j} components, along with the sampling rate. 
These components are then combined to reconstruct the complex signal. 

\textbf{Feature Extraction.} Our features are derived from the frequency domain using the Fast Fourier Transform (FFT). 
The FFT is an efficient algorithm for computing the Discrete Fourier Transform (DFT) ~\cite{gao2024parameter}, reducing computational complexity from $O(N^2)$ to $O(N \log N)$ via a divide-and-conquer strategy~\cite{swarztrauber1984fft}. 
For a discrete time-domain sequence $x[n]$ of length $N$, the DFT is defined as:
\begin{equation}
X[k] = \sum_{n=0}^{N-1} x[n] \cdot e^{-j \frac{2\pi kn}{N}}, \quad k = 0, 1, \ldots, N-1
\end{equation}
with the corresponding frequency components given by:
\begin{equation}
f_{k} = \frac{k \cdot F_{s}}{N}
\end{equation}
where $F_{s}$ denotes the sampling rate, which must satisfy the Nyquist–Shannon sampling theorem ~\cite{por2019nyquist} ($F_{s} > 2f_{\text{max}}$, Where $f_{\text{max}}$ is the maximum frequency of the signal to be collected) to avoid aliasing.  

In terms of calculation, the expression of the discrete Fourier transform is:
\begin{equation}
\mathbf{X}[k]=\sum_{n=0}^{N-1} x[n] \cdot W_{N}^{k n}, \quad W_{N}=e^{\frac{-j 2 \pi}{N}}
\end{equation}
FFT uses the divide-and-conquer method to decompose the DFT into recursive calculations of odd and even subsequences:
\begin{equation}
X[k]=\sum_{m=0}^{\frac{N}{2}-1} x[2m] \cdot W_{N/2}^{k m}+W_{N}^{k} \sum_{m=0}^{N/2-1} x[2m+1] \cdot W_{N/2}^{k m}
\end{equation}
By recursively decomposing and rotating the periodicity of the factors, repeated computations can be reduced.

The signal after FFT can also be returned to the time domain through the inverse transform:
\begin{equation}
\mathrm{x}[n]=\frac{1}{N} \sum_{k=0}^{N-1} X[k] \cdot e^{\frac{j 2 \pi k n}{N}}
\end{equation}
Applying FFT converts the collected complex time-domain signal into the frequency domain, exposing both the amplitude and phase information of each frequency component.

We evaluated three potential approaches for feature extraction:

\textbf{1. Raw Time-Domain Data.}
One straightforward option is to use the reconstructed complex signal directly. While this representation captures instantaneous magnitude and phase information, it fails to reveal the distribution of signal power across frequencies and is highly sensitive to environmental noise.

\textbf{2. Single Full-Signal FFT.}
A second option is to apply a single FFT to the entire signal segment. This method is computationally efficient and provides the finest frequency resolution. However, it is highly vulnerable to non-stationarity and spectral leakage. Because long-duration signals are seldom perfectly stationary, any transient pulse or external EM interference occurring mid-segment can introduce spurious high-frequency artifacts, thereby corrupting the spectrum.

\textbf{3. Averaged Short-Window FFT.}
The third method, which we adopt, applies an averaged short-window FFT with a two-level segmentation strategy.
First, the entire raw signal is divided into multiple independent segments, and each segment is treated as a separate training sample.
Within each segment, we further divide the signal into several non-overlapping short windows.
An FFT is then computed for each short window, and the resulting spectra are summed and averaged to produce a stable average power spectrum for that segment.
This approach effectively captures the frequency-domain characteristics of each segment while mitigating the impact of transient fluctuations within it.
Compared to a single FFT over the entire signal, it improves robustness by reducing the influence of local non-stationarities and transient noise.
The resulting Average Power Spectrum is a one-dimensional feature vector that describes the mean signal intensity across frequencies for each segment.
Although this representation provides enhanced stability and noise resilience, it still retains certain background interference, motivating the denoising step described next.

\subsection{Environmental Noise Processing}
\label{sec:Noisereduction}
The average power spectrum features extracted in the previous step still contain significant ambient noise. 
To isolate gesture-specific components, a robust denoising method is required.  

A simple technique such as spectral subtraction ~\cite{smith2021spectral} could be employed, but it often introduces artifacts and performs poorly under non-stationary noise conditions. 
To address this, we first employ adaptive signal decomposition to separate the raw spectrum into distinct modes.
Variational Mode Decomposition (VMD) ~\cite{dragomiretskiy2013variational} ~\cite{chen2022emd} is particularly well-suited for this task, as it decomposes the signal into narrowband components with well-defined center frequencies, allowing us to distinguish gesture-relevant modes from noise-dominated ones.
By applying VMD as a preprocessing step, we reduce the dominance of non-stationary noise and obtain cleaner signal components, on which spectral subtraction can then be performed more effectively and with fewer artifacts.

VMD extracts all modes simultaneously, adaptively determining their frequency bands, and is more robust and stable compared to EMD ~\cite{rilling2003empirical}.  
Each VMD mode is modeled as a narrowband amplitude–frequency modulation (AM–FM) signal:
\begin{equation}
u_{k}(t) = A_{k}(t) \cos \left( \Phi_{k}(t) \right),
\end{equation}
where $A_{k}(t)$ is a slowly varying amplitude $(A_{k}(t) \geq 0)$ and $\Phi_{k}(t)$ is the instantaneous phase. $u_{k}(t)$ denotes the $k$-th mode function to be estimated.
In the frequency domain, each mode is assumed to be narrowband, with its energy concentrated near its center frequency.  
The optimization objective of VMD is to minimize the bandwidth of all modes while ensuring accurate reconstruction of the original signal:
\begin{equation}
\min_{\{u_{k}\},\{w_{k}\}} \left\{ \sum_{k} \left\| \partial_{t} \left[ \left( \delta(t) + \frac{j}{\pi t} \right) \ast u_{k}(t) \right] e^{-j w_{k} t} \right\|_{2}^{2} \right\},
\end{equation}
where $u_{k}(t)$ is the $k$-th mode, and $w_{k}$ its center frequency. 
The operator $\partial_{t}$ represents the temporal derivative, emphasizing high-frequency variations. 
$\delta(t)$ is the Dirac delta function, and $\tfrac{j}{\pi t}$ is the kernel of the Hilbert transform; together $\delta(t) + \tfrac{j}{\pi t}$ is the time-domain representation of the analytic signal operator. 
The convolution $\left( \delta(t) + \tfrac{j}{\pi t} \right) \ast u_{k}(t)$ therefore constructs the analytic signal of $u_{k}(t)$. 
The exponential term $e^{-j w_{k} t}$ demodulates the signal to baseband around the center frequency $w_{k}$. 
Finally, $\|\cdot\|_{2}^{2}$ computes the squared $\ell_{2}$-norm, which corresponds to the total bandwidth energy of each mode after demodulation. The above formula also satisfies:
\begin{equation}
\sum_{k=1}^{K} u_{k}(t) = f(t),
\end{equation}
where $f(t)$ is the original signal.  

Applying VMD requires setting key hyperparameters: the number of modes $K$, the penalty factor $\alpha$, and initialization values for modes, center frequencies, and Lagrange multipliers. 
The optimization proceeds iteratively through:  
(1) modal updates via Wiener filtering in the frequency domain,  
(2) center frequency updates by computing the spectral centroid,  
(3) reconstruction constraints enforced through Lagrange multipliers, and  
(4) convergence checks, terminating when mode updates fall below a threshold.  

In our workflow, both the gesture-contaminated spectrum and an independently recorded noise spectrum are first decomposed using VMD with the same parameters.
This produces two sets of mode components that are frequency-aligned across signal and noise.
We then compare the corresponding modes from the gesture signal and the noise signal, and apply spectral subtraction at the mode level.
In this way, structured noise such as power-line hum and broadband interference can be more accurately removed, while gesture-related spectral patterns are preserved.
This two-step process—VMD decomposition followed by mode-wise spectral subtraction—provides a cleaner representation of gesture features and reduces the artifacts typically introduced by direct spectral subtraction.

For analysis, we retain three versions of the signal: the \textit{original} spectrum (gesture signal with noise), the \textit{noise} spectrum (only noise), and the \textit{denoised} spectrum (gesture signal after mode-wise spectral subtraction using the corresponding noise modes). The final denoised spectrum is stored in an \texttt{.npz} file and serves as the input feature for model training.

\subsection{Gesture Recognition}
The VMD-based denoising step provides our final feature representation: a clean average power spectrum (APS) ~\cite{mohammadzadeh2022robust}. 
We segment the raw recordings into segments, each corresponding to a single gesture sample. 
Every segment is processed through the full pipeline (signal reconstruction, short-window FFT averaging, and VMD-based denoising) to produce one APS vector. 
By collecting a sufficient number of labeled APS vectors for each gesture type, we construct the dataset for model training.  

Given that our feature engineering pipeline already produces discriminative gesture features, it is unnecessary to employ computationally intensive deep learning models. 
Instead, we adopt a lightweight yet effective Random Forest classifier~\cite{breiman2001random}. 
Random Forest is an ensemble method that trains multiple decision trees on bootstrap samples of the data, with each tree considering a random subset of features at each split. 
The final prediction is obtained through majority voting across all trees. 
This approach improves classification accuracy, stability, and robustness to overfitting, while keeping computational cost low. We also compared several classification models in subsequent experiments. The experiments showed that the random forest model achieved better classification results. For details, please refer to Section ~\ref{sec:classification model}.

In our implementation, the denoised APS dataset is randomly split into training and testing sets with an 80/20 ratio. 
The Random Forest is trained on the training set and evaluated on the unseen testing set. 

%% file: prototype.tex
\section{Prototype}
\begin{figure}[tbp]
  \centering
  \begin{subfigure}{0.32\textwidth}
    \centering
    \includegraphics[width=\linewidth]{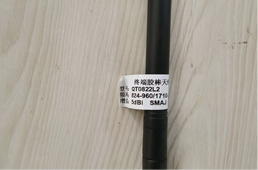} % Replace with actual filename
    \caption{Antenna with $5$ dB gain}
    \label{Antenna}
  \end{subfigure}
  \hfill
  \begin{subfigure}{0.32\textwidth}
    \centering
    \includegraphics[width=\linewidth]{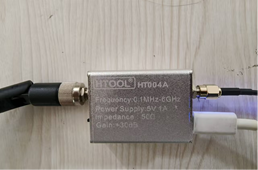} % Replace with actual filename
    \caption{Signal amplifier with $30$ dB gain}
    \label{Amplifier}
  \end{subfigure}
  \hfill
  \begin{subfigure}{0.32\textwidth}
    \centering
    \includegraphics[width=\linewidth]{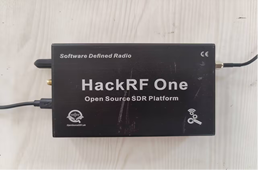} % Replace with actual filename
    \caption{HackRF One (SDR)}
    \label{HackRF}
  \end{subfigure}
  \caption{Hardware components of the \textit{EMGesture} prototype.}
  \label{Prototype}
\end{figure}

This section presents the hardware implementation of \textit{EMGesture}. 
Our prototype consists of four main components: an antenna, a signal amplifier, the HackRF software-defined radio (SDR) platform, and a computer (analyzer) for data processing and analysis. 
Some components of the system prototype is illustrated in Fig.~\ref{Prototype}.

The workflow proceeds as follows: (1) the antenna captures gesture-perturbed electromagnetic signals and converts them into analog voltage signals; 
(2) the universal signal amplifier boosts the signal strength while reducing interference; 
(3) the amplified signals are digitized by the HackRF SDR at a customized sampling rate and streamed to the analyzer; 
(4) the analyzer executes the full processing pipeline, including feature extraction, denoising, and classification.  

A key advantage of our approach is that the classification pipeline—based on VMD and Random Forest is computationally lightweight, avoiding the high resource demands of deep learning. 
This design enables a clear path toward future miniaturization: the entire prototype, including a compact antenna, amplifier, sampler, and microcontroller running the classifier, could be seamlessly integrated into the housing of a wireless charger. 
Such integration would yield a fully self-contained and unobtrusive interactive device, requiring no external hardware components.

%% file: evaluation.tex
\section{Evaluation}
\subsection{Experiment Setup}
Unless otherwise noted, all experiments were conducted under the following configuration. 
Interaction hardware: a Belkin 15W magnetic wireless charger with an iPhone 14 Plus actively charging. 
Sensing hardware: a receiving antenna (5 dB gain) positioned approximately 25 cm above the charger, connected to an HT004A signal amplifier (30 dB gain). 
The amplified signal was then fed into a HackRF SDR, which digitized the data at a 20 MHz sampling rate and streamed it to a analyzer for processing.

\textbf{Data Collection Protocol.}  
After powering on the charger and placing the phone, we waited for the electromagnetic (EM) field to stabilize. 
Gestures were performed approximately 12 cm above the charger (between the phone and antenna). 
We defined nine classes: one baseline “no gesture” (active charging) and eight distinct hand gestures (Gesture 1, Gesture 2, Gesture 3, Gesture 4, Hand-spreading, Gesture OK, Gesture 8, and Fist). 
Each gesture was recorded for about 30 seconds, from which we used the 2nd to 25th second (23 seconds) to exclude transient noise at the start and end.

\textbf{Dataset Generation.}  
The effective 23-second recording for each gesture was segmented into 0.5-second windows, each further divided into 0.01-second sub-windows for power spectrum accumulation. 
This resulted in 46 data samples per gesture class (23/0.5). 
Given a 20 MHz sampling rate and a 0.01-second sub-window, each FFT produced 200,000 points, yielding a feature vector of length 200,000. 
Thus, the complete dataset contained $46 \times 9 = 414$ samples, each represented by a 200,000-dimensional vector.

\textbf{Data Preprocessing.}  
To enable spectral subtraction, we also collected an ambient noise profile by disconnecting the wireless charger (while keeping the phone and antenna in place). 
This noise data was processed through the same pipeline as the gesture data (segmentation, sub-windowing, and power spectrum computation) to obtain a stable noise profile. 
In the denoising stage, spectral subtraction was applied bin-by-bin between the noise profile and the corresponding gesture spectrum, producing a clean feature vector that preserved gesture-specific perturbations. 
These denoised features were then used to train the Random Forest classifier, with performance evaluated using classification accuracy and a confusion matrix.

\subsection{Micro-benchmark}
\subsubsection{Choosing Data Collection Equipment}
Before finalizing our hardware pipeline (antenna and SDR), we evaluated two alternative sensing devices and compared their performance with our chosen approach.

\begin{figure}[tbp]
  \centering
  \includegraphics[width=0.4\linewidth]{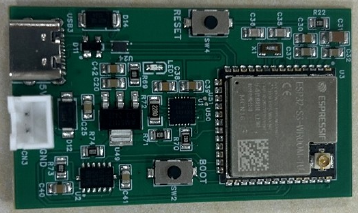} 
  \caption{Custom ESP32-based electromagnetic signal collection board}
  \label{Board}
\end{figure}

\textbf{Alternative 1: Custom ESP32 Magnetic Field Sensor.}  
We first designed and prototyped a custom ESP32-S3 development board equipped with a DRV425 fluxgate sensor, which outputs an analog voltage proportional to magnetic field strength (Fig.~\ref{Board}). The analog signal was connected to ADC1 channel 4 (GPIO4) of the ESP32-S3, and the digitized data was transmitted via UDP to a host computer.  

\begin{figure}[tbp]
  \centering
  \begin{subfigure}{0.48\textwidth}
    \centering
    \includegraphics[width=\linewidth]{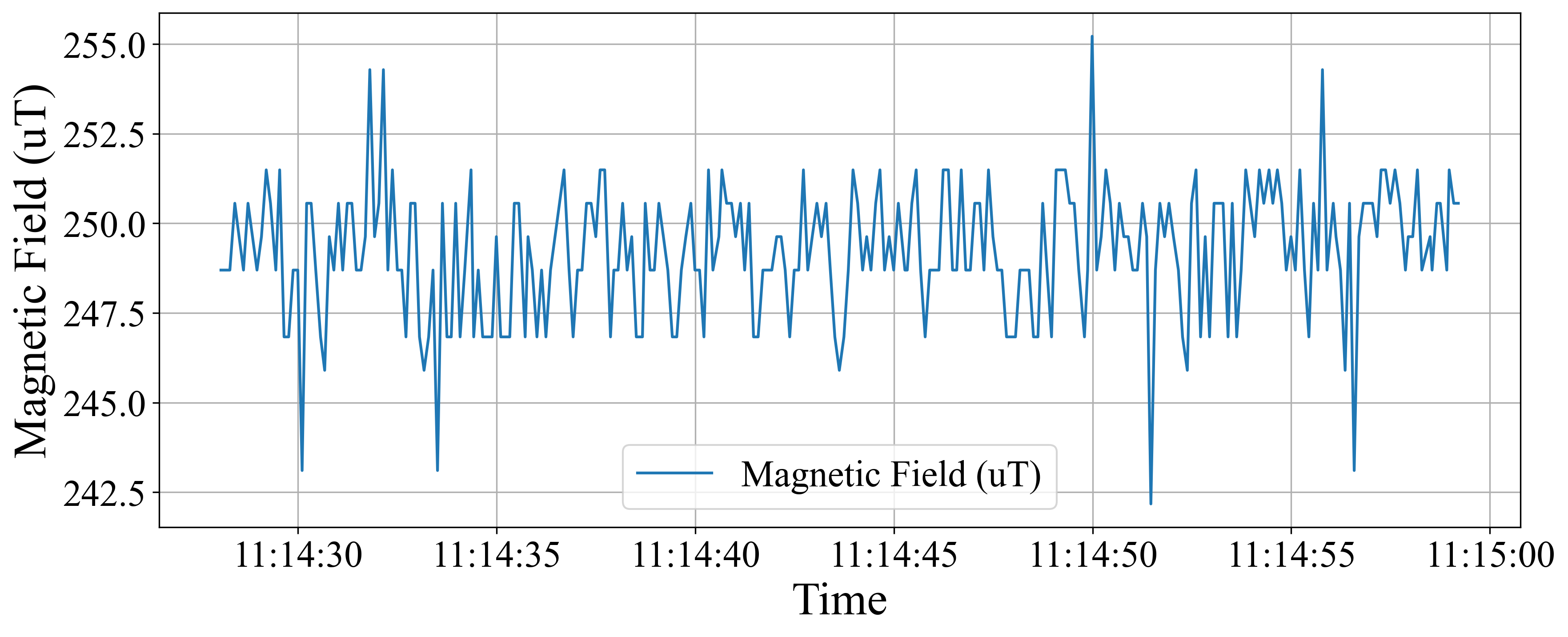} 
    \caption{Time-domain signal}
    \label{BoardTime}
  \end{subfigure}
  \hfill
  \begin{subfigure}{0.48\textwidth}
    \centering
    \includegraphics[width=\linewidth]{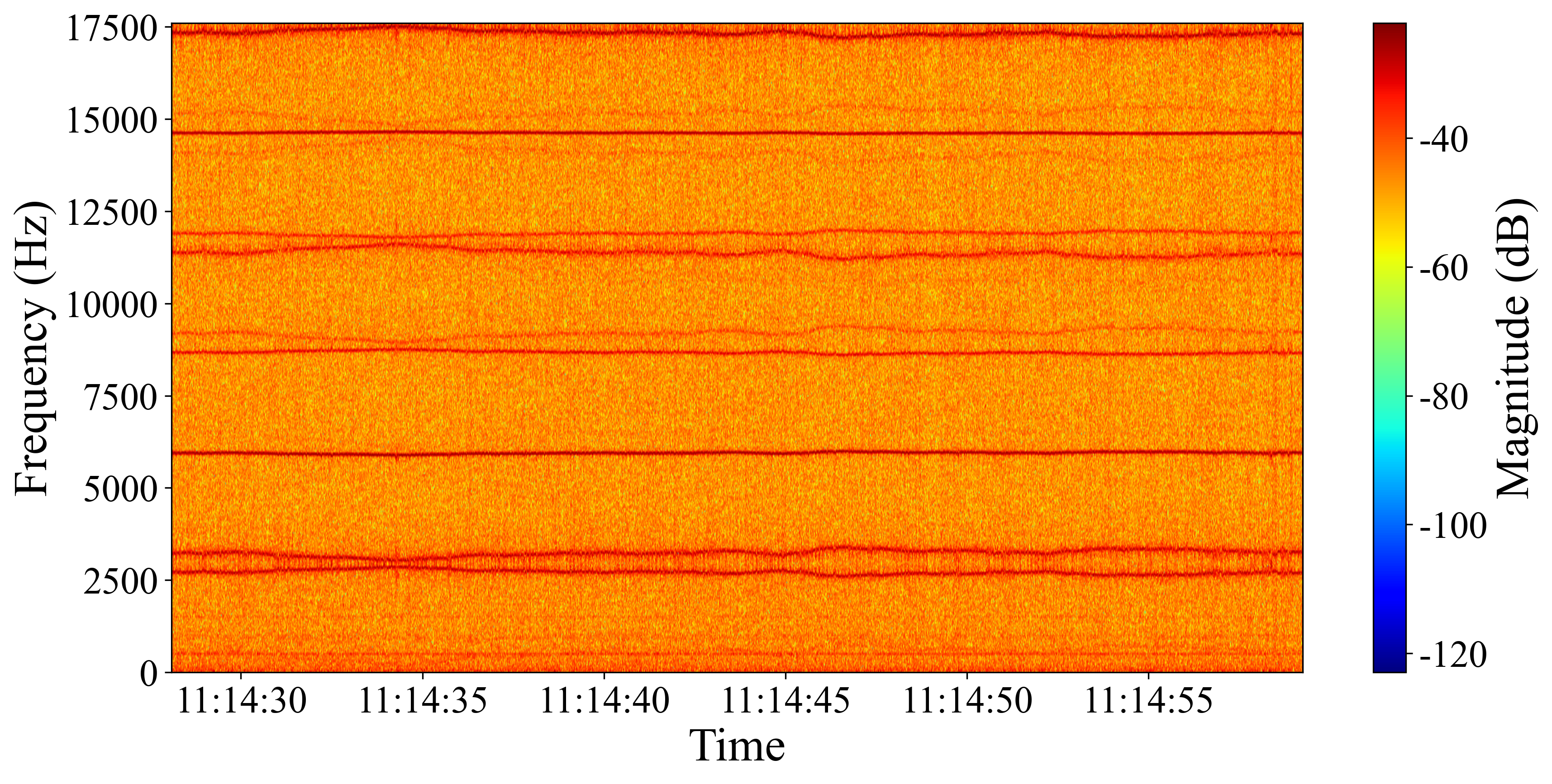} 
    \caption{Spectrogram}
    \label{BoardFrequency}
  \end{subfigure}
  \caption{Visualization of signals collected using the ESP32 development board}
  \label{SignalBoard}
\end{figure}

We collected data using this board and visualized it both in the time domain (Fig.~\ref{BoardTime}) and as a spectrogram (Fig.~\ref{BoardFrequency}). These figures covered an 8-second capture: four seconds of “no gesture” baseline followed by four seconds of a “left–right swing” gesture. As shown, no significant differences were observed between the two phases, and classification performance was no better than random guessing. We attribute this failure to the ESP32’s limited sampling rate (about 33 kHz), which, according to the Nyquist theorem, restricts capture to frequencies below 16.5 kHz. This suggests that gesture-related perturbations mainly reside in higher-frequency bands beyond the ESP32’s sensing range, motivating our decision to adopt a wideband SDR-based solution.

\begin{figure}[tbp]
  \centering
  \includegraphics[width=0.7\linewidth]{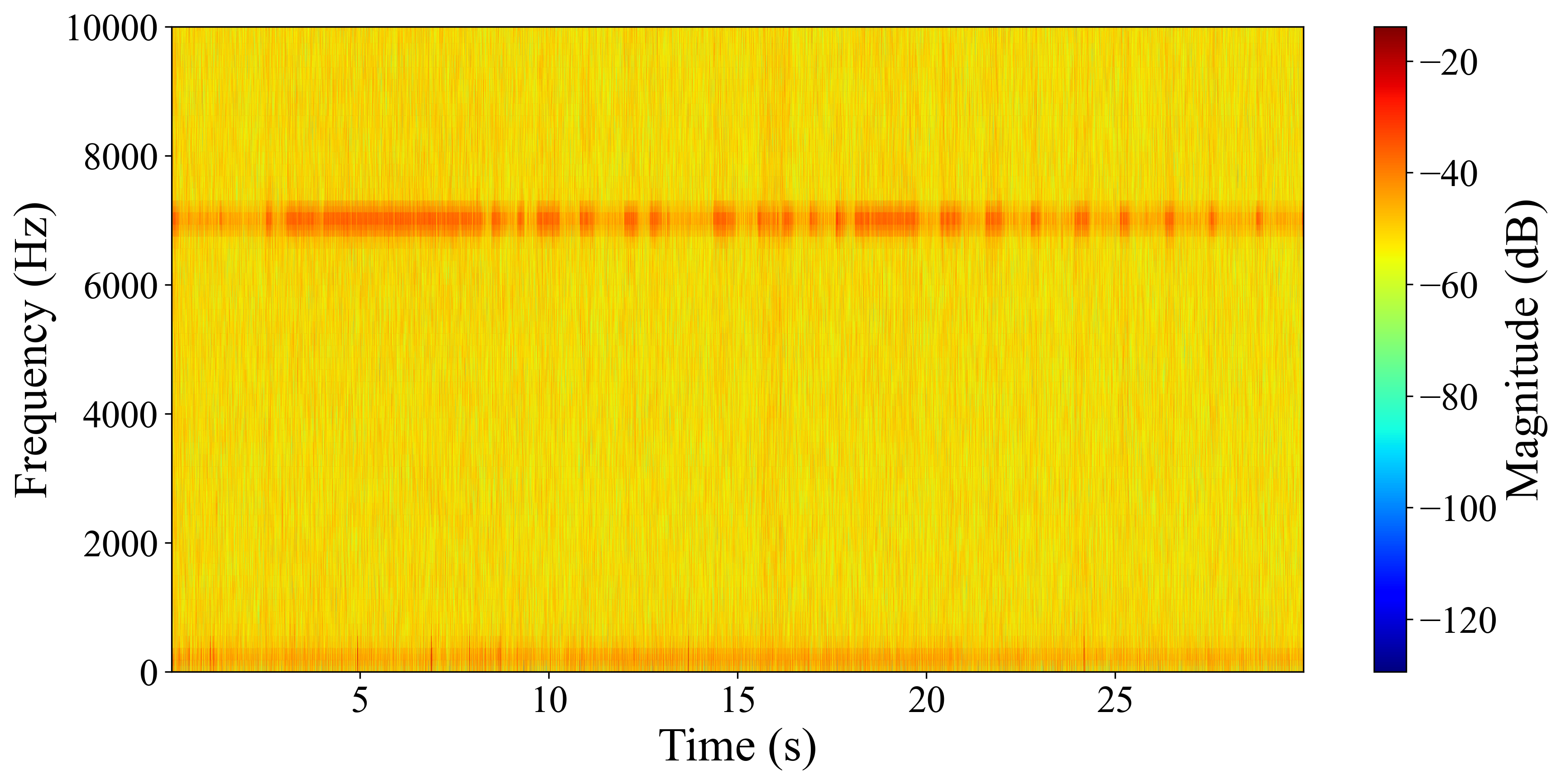} 
  \caption{Spectrum of EM signals captured using a sound card}
  \label{SoundcardFrequency}
\end{figure}

\textbf{Alternative 2: Sound Card.}  
Inspired by AUDIOSENSE~\cite{li2023audiosense}, which demonstrated measuring EM waves via audio sensing, we tested a standard sound card (192 kHz sampling rate) as a low-cost acquisition tool. To further validate feasibility, we artificially generated stable EM signals by modulating CPU duty cycles, creating periodic loads on the phone. Fig.~\ref{SoundcardFrequency} shows an example spectrum obtained using this method. Although the sound card successfully captured low-frequency EM signals, its effective sensing range was limited to approximately 1 cm, making it unsuitable for practical gesture recognition tasks. In subsequent work (see Sec.~\ref{sec:feedback}), we leveraged this setup to test the wireless charger’s negative feedback mechanism for active modulation.

\textbf{Final Choice: Wideband Antenna with SDR.}  
The above two approaches failed due to insufficient bandwidth (ESP32) and insufficient range (sound card). We therefore adopted a wideband antenna (5 dB gain) connected to a HackRF SDR. This setup captures free-space EM waves over a wide frequency range and at longer distances.  

\begin{figure}[tbp]
  \centering
  \includegraphics[width=0.7\linewidth]{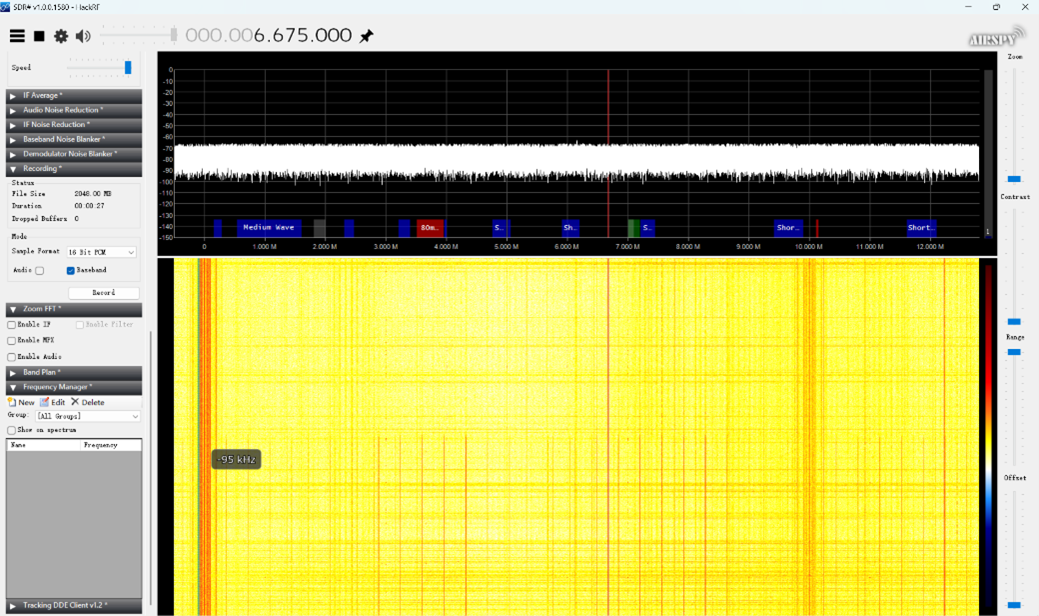} 
  \caption{Spectrum response to a hand-spreading gesture}
  \label{AntennaHand}
\end{figure}

Our results confirmed that the antenna–SDR pipeline robustly detected wireless charger emissions up to 2 m away, despite the modest antenna gain. Moreover, the captured signals were highly responsive to hand gestures. As shown in Fig.~\ref{AntennaHand}, a hand-spreading gesture caused clear attenuation across multiple frequency bands—sensitivity absent in the ESP32 and sound card methods. These results demonstrate that the antenna–SDR pipeline provides the necessary bandwidth, sensitivity, and range for our task. Consequently, this configuration was adopted for all subsequent experiments.

\subsubsection{Choosing a Classification Model}
\label{sec:classification model}
To identify the most suitable lightweight classifier for our task, we compared three widely used models: Random Forest (RF) ~\cite{hu2023review}, K-Nearest Neighbors (KNN) ~\cite{zhang2021challenges}, and Support Vector Machine (SVM) ~\cite{valkenborg2023support}. Given the high dimensionality of our features (200,000 dimensions), we applied PCA-based ~\cite{dai2025robust} dimensionality reduction before training the KNN and SVM to mitigate the curse of dimensionality.  

\begin{figure}[tbp]
  \centering
  \begin{subfigure}{0.31\textwidth}
    \centering
    \includegraphics[width=\linewidth]{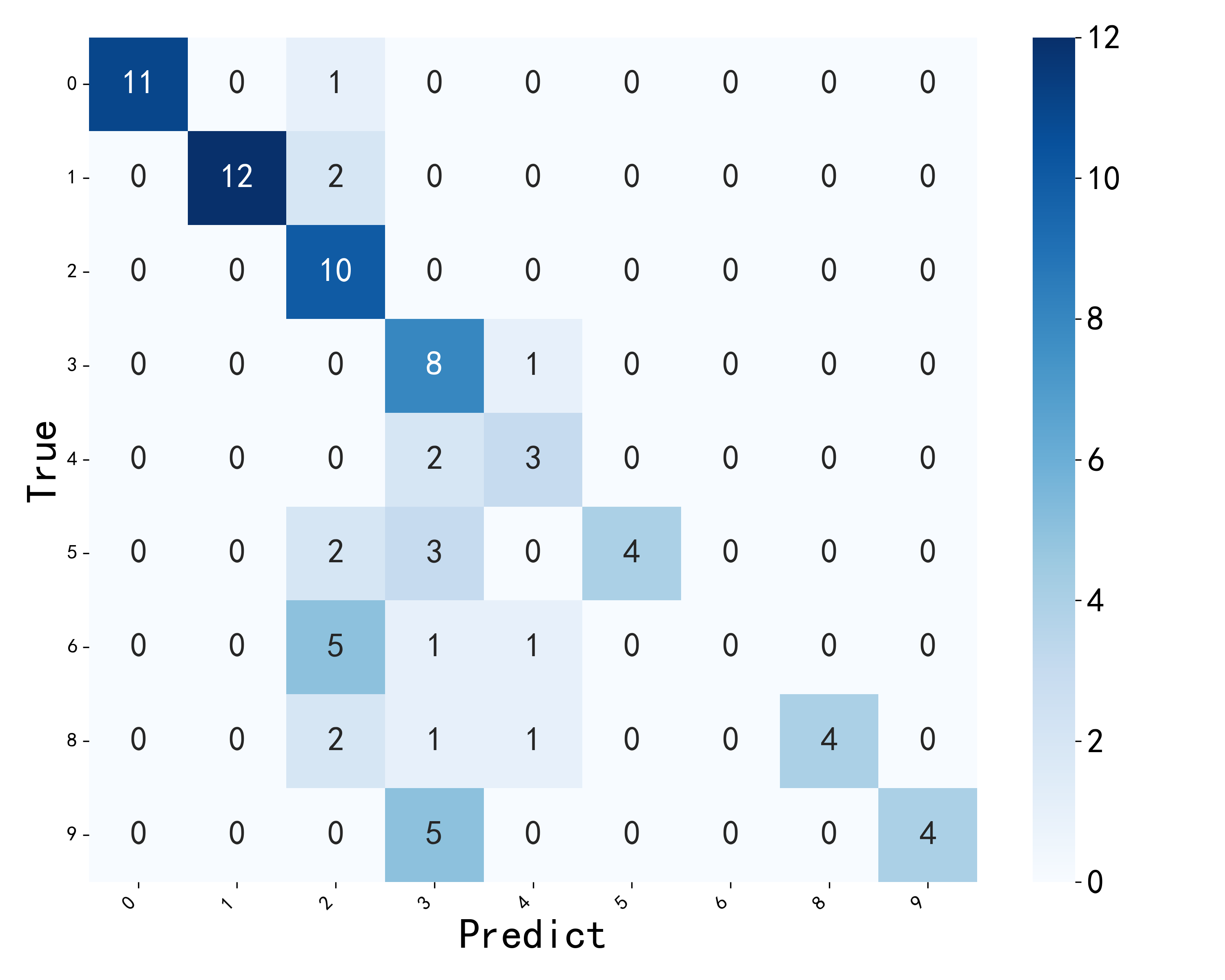} % Replace with actual filename
    \caption{KNN}
    \label{KNN}
  \end{subfigure}
  \hfill
  \begin{subfigure}{0.31\textwidth}
    \centering
    \includegraphics[width=\linewidth]{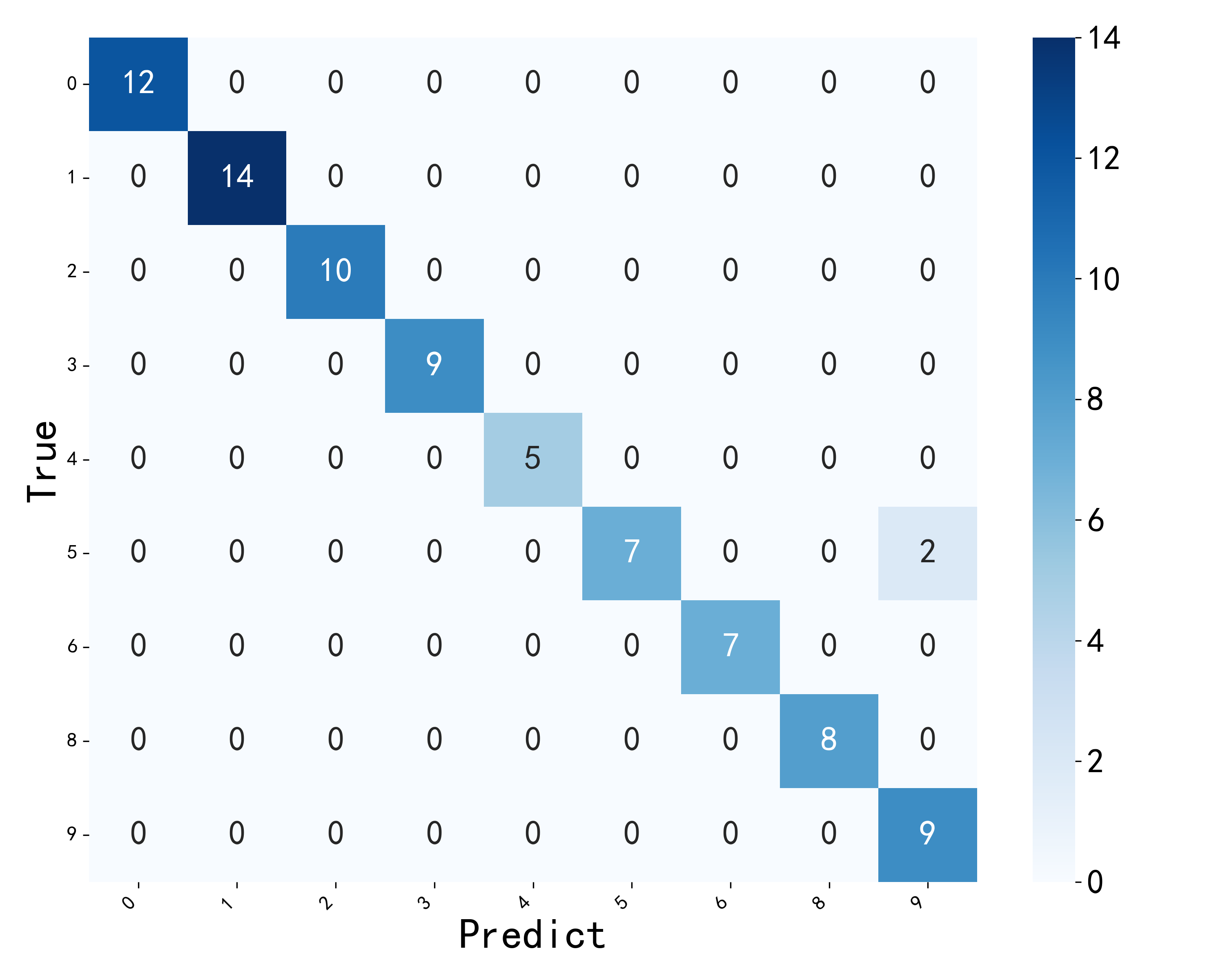} % Replace with actual filename
    \caption{Random Forest}
    \label{randomforest}
  \end{subfigure}
  \hfill
  \begin{subfigure}{0.31\textwidth}
    \centering
    \includegraphics[width=\linewidth]{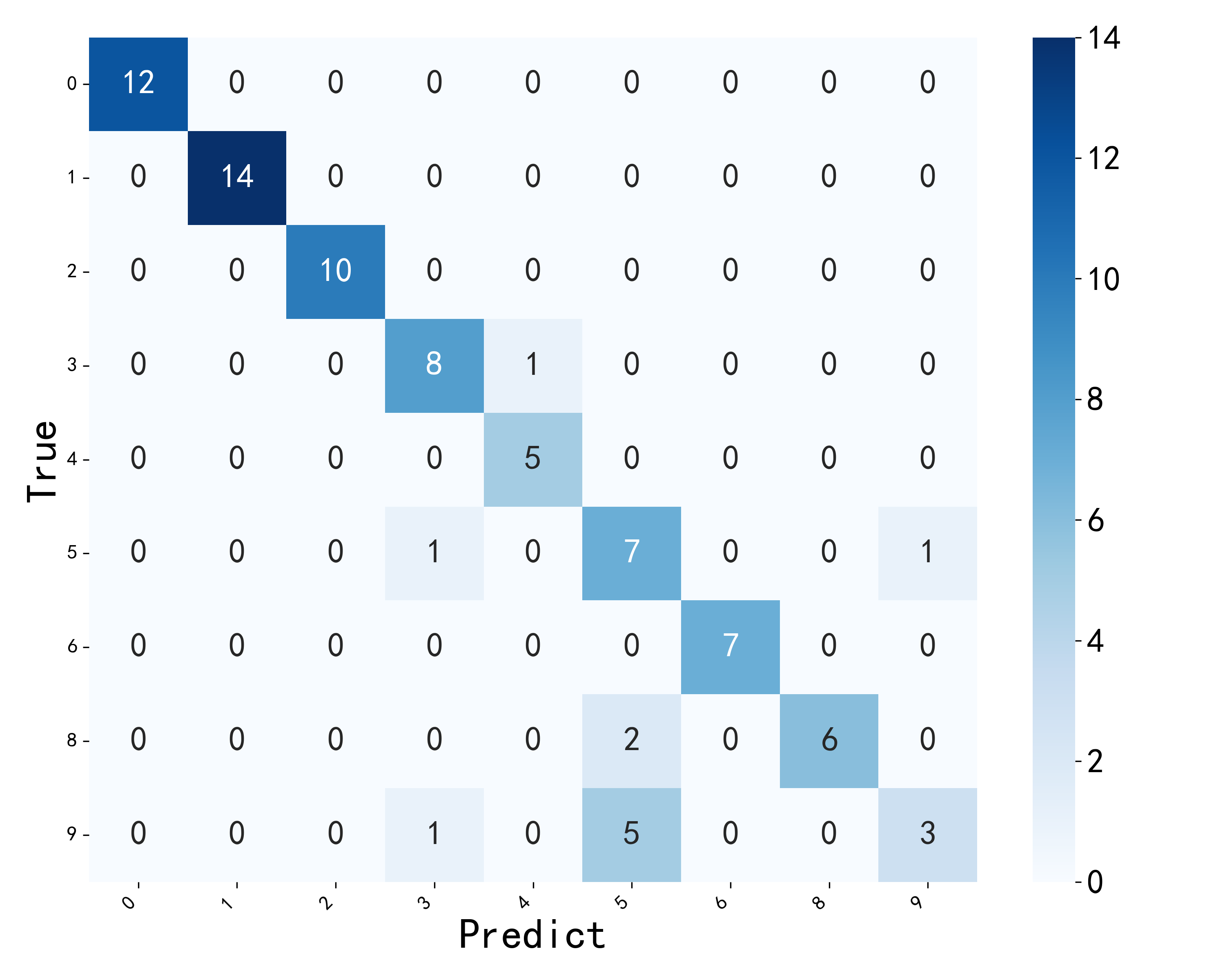} % Replace with actual filename
    \caption{SVM}
    \label{SVM}
  \end{subfigure}
  \caption{Confusion matrices of different classifiers}
  \label{classifiers}
\end{figure}

The classification results are summarized in Table~\ref{tab:model-acc}. Despite PCA preprocessing, Random Forest achieved the highest accuracy (97.59\%), significantly outperforming SVM (86.75\%) and KNN (67.47\%). The corresponding confusion matrices in Fig.~\ref{classifiers} further illustrate the superior performance of RF.  

This advantage can be attributed to the ensemble nature of Random Forest. By aggregating multiple decision trees, RF effectively captures nonlinear relationships and complex feature interactions. Moreover, decision trees inherently perform feature selection during the splitting process, making RF naturally robust to high-dimensional input without requiring additional dimensionality reduction. In contrast, PCA only provides a linear projection, which cannot fully resolve the nonlinear separability in our data. Consequently, SVM performance was limited, as it is highly sensitive to kernel choice and parameter tuning, particularly in high-dimensional spaces with relatively small training sets. KNN fared even worse, as distance-based methods remain susceptible to the curse of dimensionality even after PCA, making it difficult to compute meaningful distances and establish accurate decision boundaries.  

\begin{table}[tbp]
\centering
\caption{Classification Accuracy of Different Models}
\label{tab:model-acc}
\color{black}
\begin{tabular}{l r}
\toprule
\textbf{Model} & \textbf{Classification Accuracy} \\
\midrule
Random Forest & 97.59\% \\
Support Vector Machine (SVM) & 86.75\% \\
k-Nearest Neighbors (KNN) & 67.47\% \\
\bottomrule
\end{tabular}
\end{table}

\subsection{Overall Performance}
We next evaluate the overall performance and robustness of our approach by analyzing its adaptability to diverse users, mobile phones, and wireless chargers. 

\begin{figure}[tbp]
  \centering
  \includegraphics[width=0.4\linewidth]{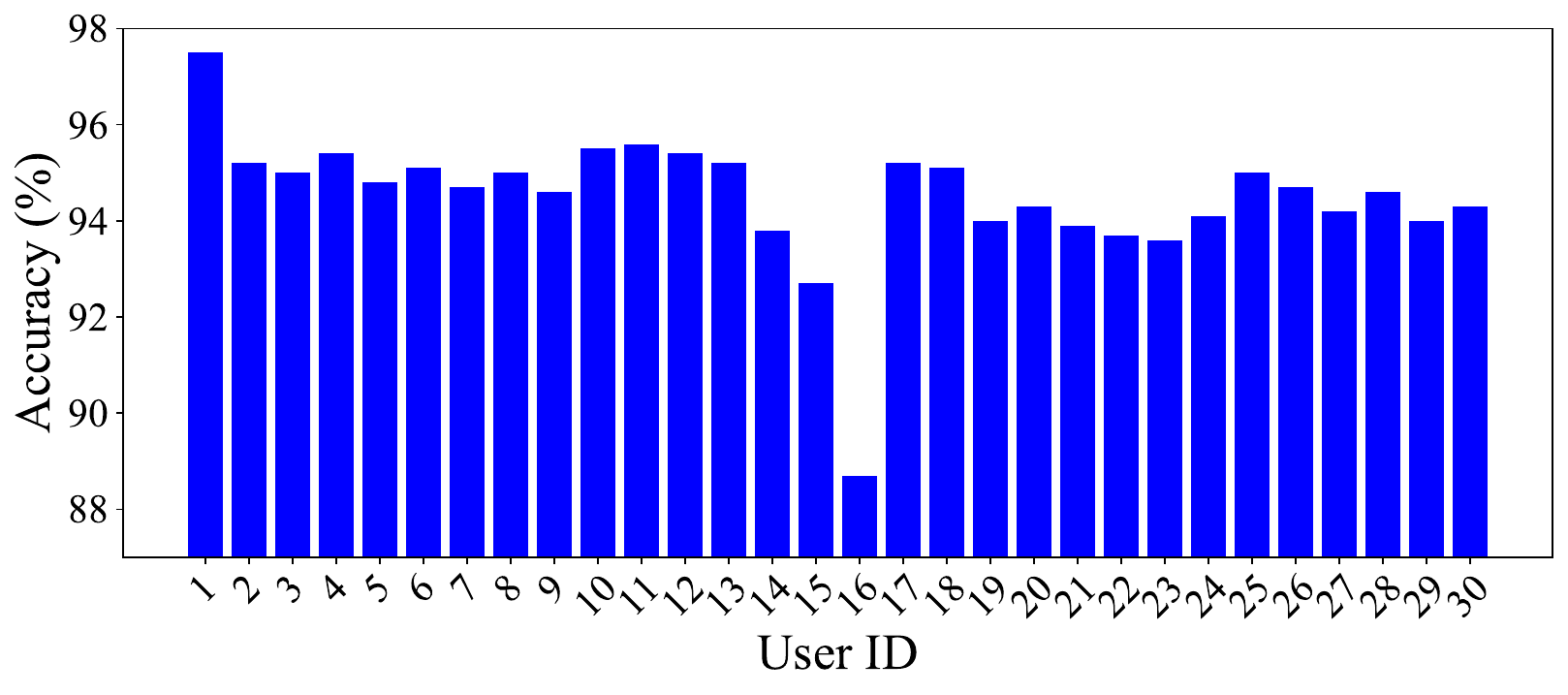} 
  \caption{Gesture recognition accuracy of 30 users using iPhone 14 Plus on a Belkin charger.}
  \label{DifferentPerson}
\end{figure}

\begin{figure}[tbp]
  \centering
  \includegraphics[width=0.4\linewidth]{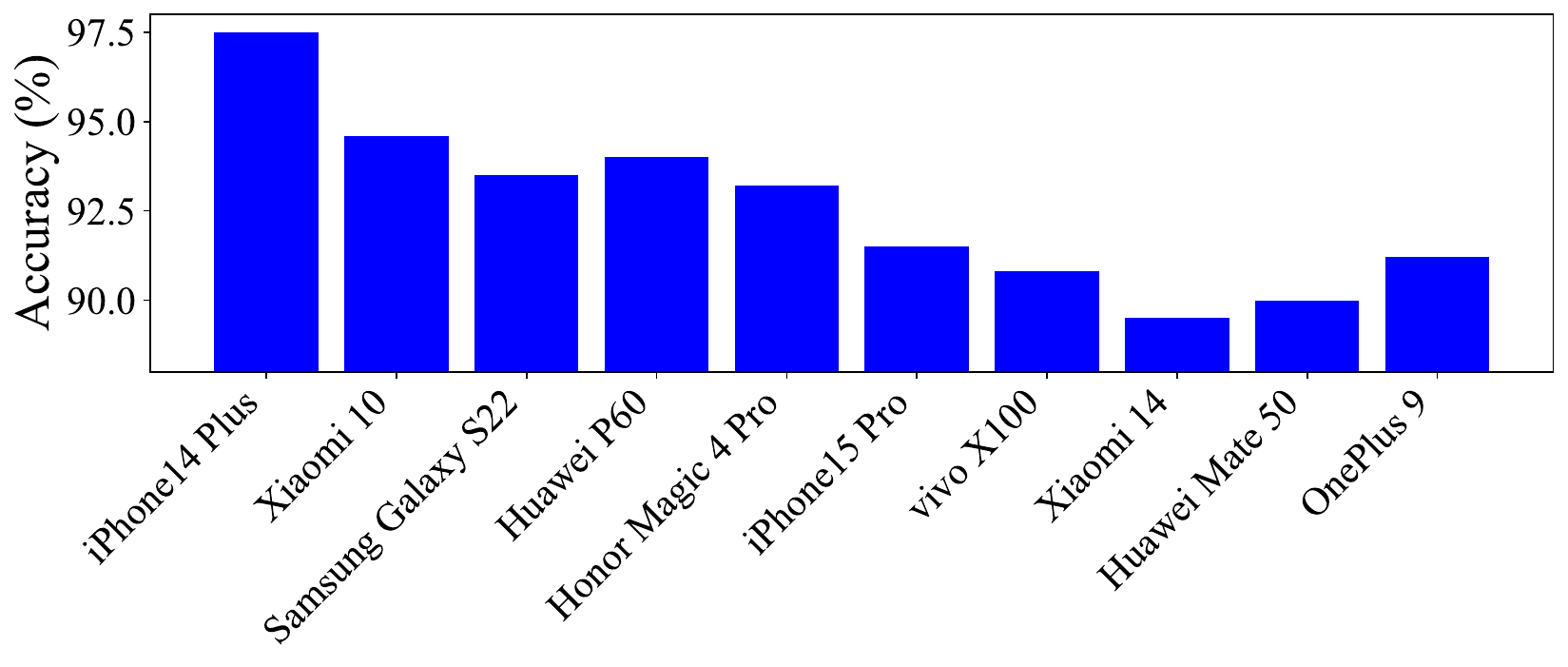} 
  \caption{Gesture recognition accuracy across different phone models on a Belkin charger.}
  \label{DifferentPhone}
\end{figure}

\begin{figure}[tbp]
  \centering
  \includegraphics[width=0.4\linewidth]{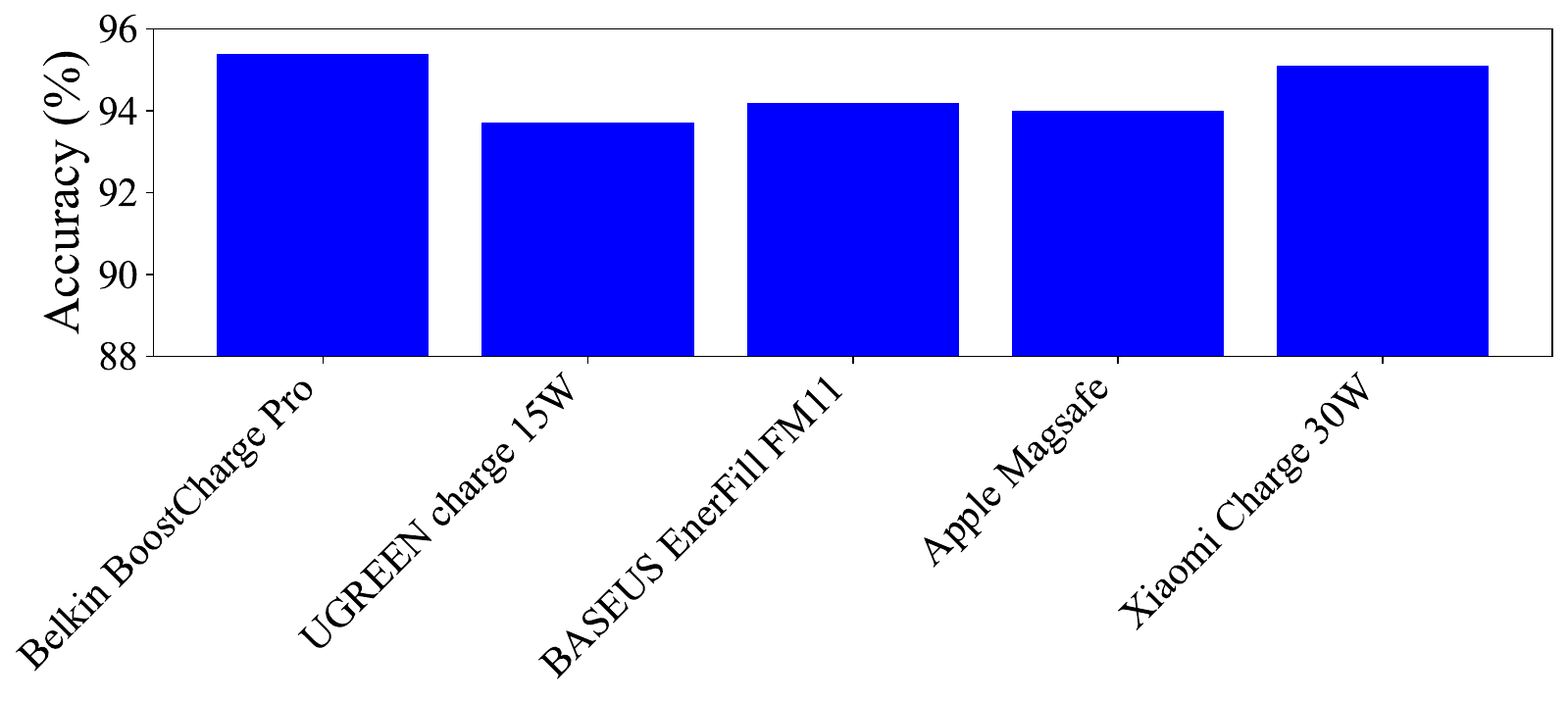} 
  \caption{Gesture recognition accuracy of iPhone 14 Plus on different wireless chargers.}
  \label{DifferentCharger}
\end{figure}

To assess robustness in real-world scenarios, we designed three independent experiments to test the system under variability in: (1) users, (2) phone models, and (3) wireless chargers. First, we collected gesture data from 30 distinct users using the baseline hardware. As shown in Fig.~\ref{DifferentPerson}, the system consistently achieved high accuracy, ranging from 88.7\% to 97.5\%, with an average of 94.50\%. Second, we tested the system on 10 different mobile phone models. As shown in Fig.~\ref{DifferentPhone}, performance exhibited slightly greater variance due to device heterogeneity, with accuracy ranging from 89.5\% to 97.5\% and an average of 92.58\%. Third, we evaluated performance across 10 wireless charger models. Results in Fig.~\ref{DifferentCharger} demonstrate that the system remained highly stable, with accuracy tightly distributed between 93.7\% and 95.4\%. 

The slight decrease observed when testing across diverse phone models can be attributed to hardware differences and variations in EM field patterns. Nevertheless, maintaining an average accuracy above 92\% across all three studies highlights that our method is both robust and adaptable, ensuring reliable performance under diverse user, device, and hardware conditions.

\subsection{Impact of Environment}
According to ~\cite{balanis2012advanced}, the intensity of an electromagnetic (EM) wave decays as it propagates through a conductive medium, which can be expressed as:  

\begin{equation}
	\delta = \sqrt{\frac{2}{\mu _{0}\varepsilon _{0}}}\frac{1}{\omega \left(\sqrt{1+\left(\frac{\sigma}{\omega\varepsilon _{0}}-1\right)^{1/2}}\right)}\simeq \sqrt{\frac{2}{\mu _{0}\sigma \omega}},
\end{equation}
where $\delta$ is the skin depth, defined as the depth below the conductor’s surface at which the current density decreases to $1/e$ of its value at the surface (approximately $0.37$). $\varepsilon_0$ denotes the permittivity of free space, $\mu_0$ the permeability of free space, $\sigma$ the electrical conductivity, and $\omega=2\pi f$, with $f$ representing the EM signal frequency.  

\begin{figure}[tbp]
  \centering
  \includegraphics[width=0.5\linewidth]{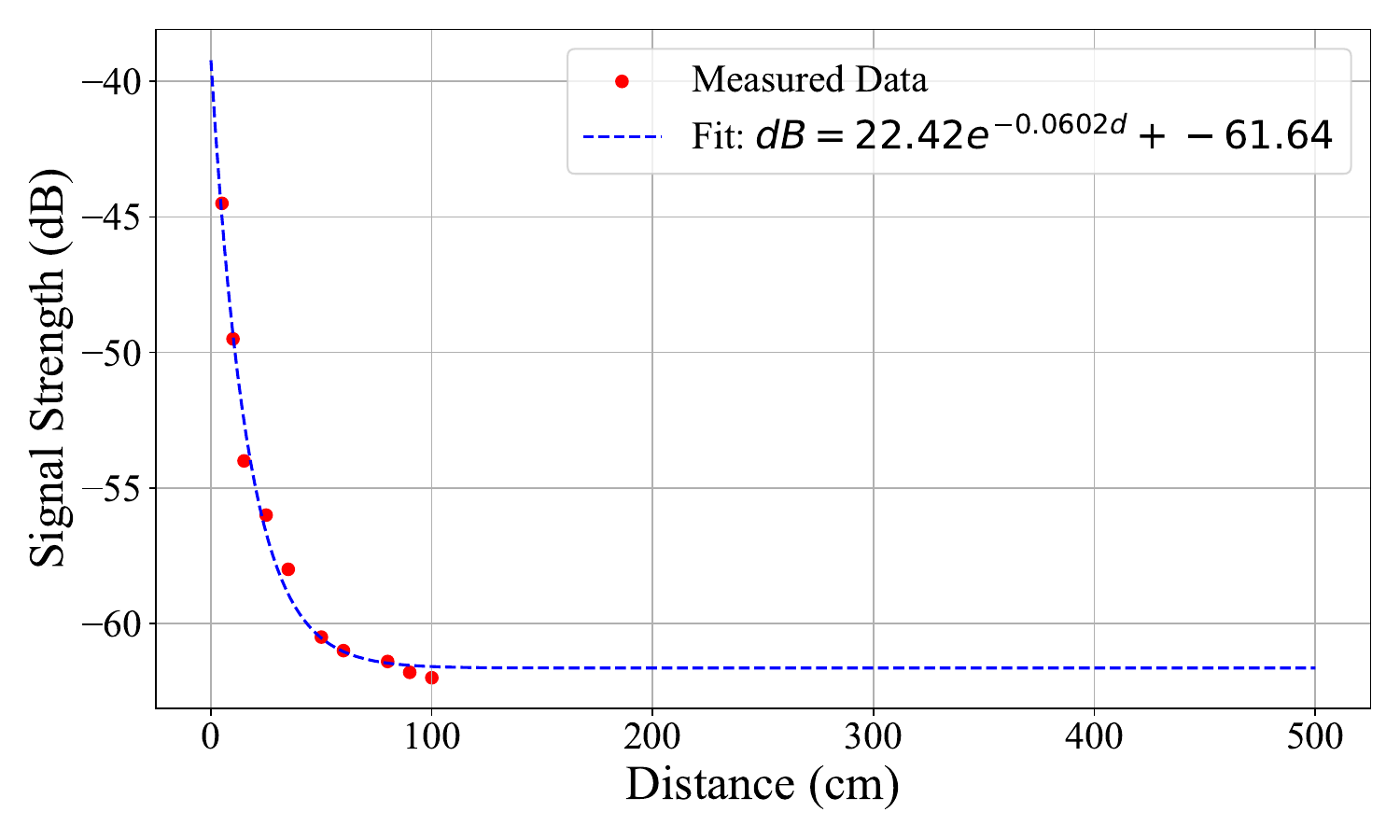} 
  \caption{Attenuation of EM signals with increasing distance.}
  \label{fig:EMdistance}
\end{figure}

This principle implies that EM signals attenuate exponentially with distance. Such rapid decay is advantageous for our system, as it ensures that the sensor primarily captures near-field perturbations, while distant environmental noise has minimal influence. To validate this property, we conducted an experiment (Fig.~\ref{fig:EMdistance}) by placing a magnetic probe at varying distances from a strong EM source and recording the resulting intensity. The results confirmed an exponential decay pattern. This demonstrates that the captured signals are highly localized, supporting our assumption that distant EM noise has negligible impact on gesture recognition.  

\begin{figure}[tbp]
  \centering
  \includegraphics[width=0.5\linewidth]{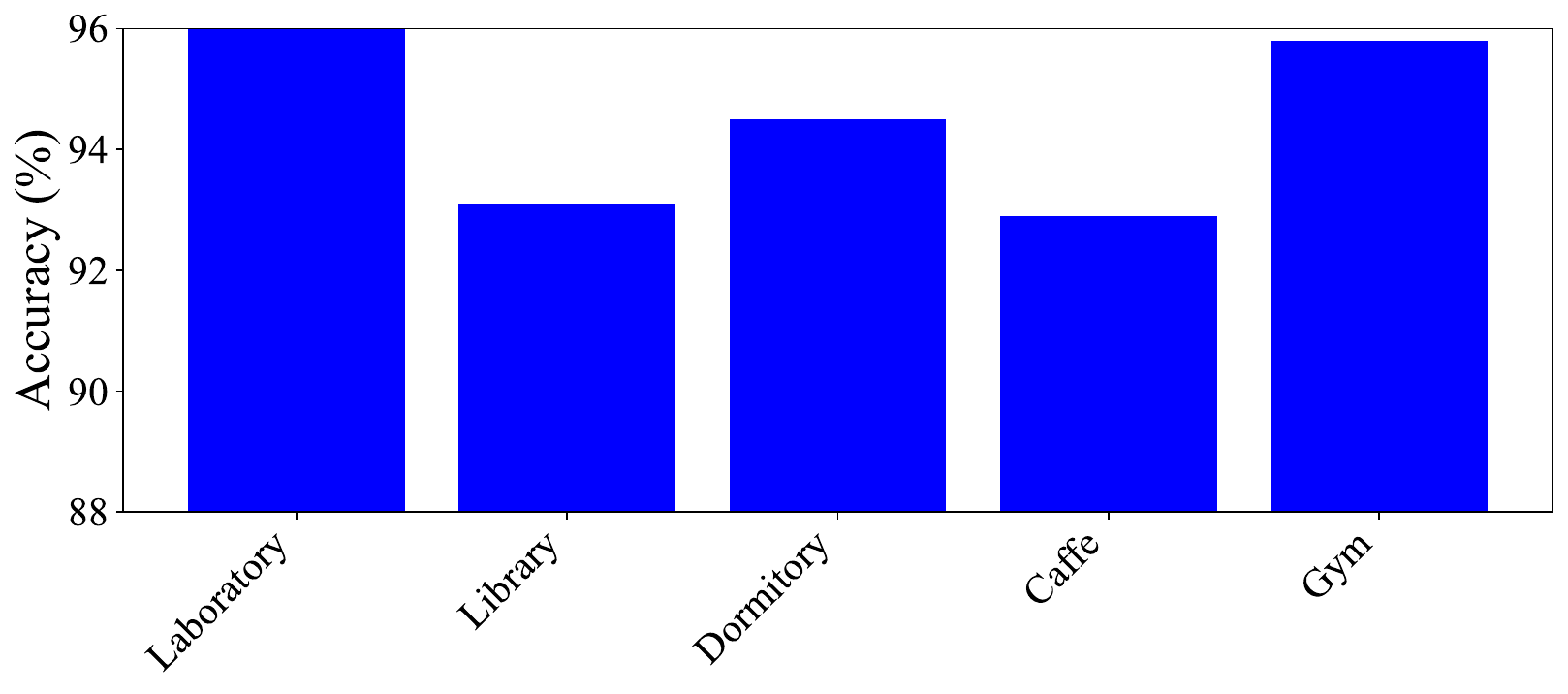} 
  \caption{Gesture recognition accuracy across different environments using iPhone 14 Plus on a Belkin charger.}
  \label{fig:DifferentEnvironment}
\end{figure}

To further assess robustness against environmental EM interference, we tested our system across five diverse real-world settings: a gymnasium, a library, a cafe, a dormitory, and a laboratory (containing numerous computers and electronic devices). The results, shown in Fig.~\ref{fig:DifferentEnvironment}, indicate that the system consistently achieved an average gesture recognition accuracy of 94\% across all environments. This confirms that our approach is resilient to environmental variability and largely unaffected by background EM interference.  

\subsection{Validation of Spectral Subtraction Denoising} 
To evaluate the effectiveness of our denoising strategy, we first visualize the signals before and after processing.  

\begin{figure}[tbp]
    \centering
    \includegraphics[width=\linewidth]{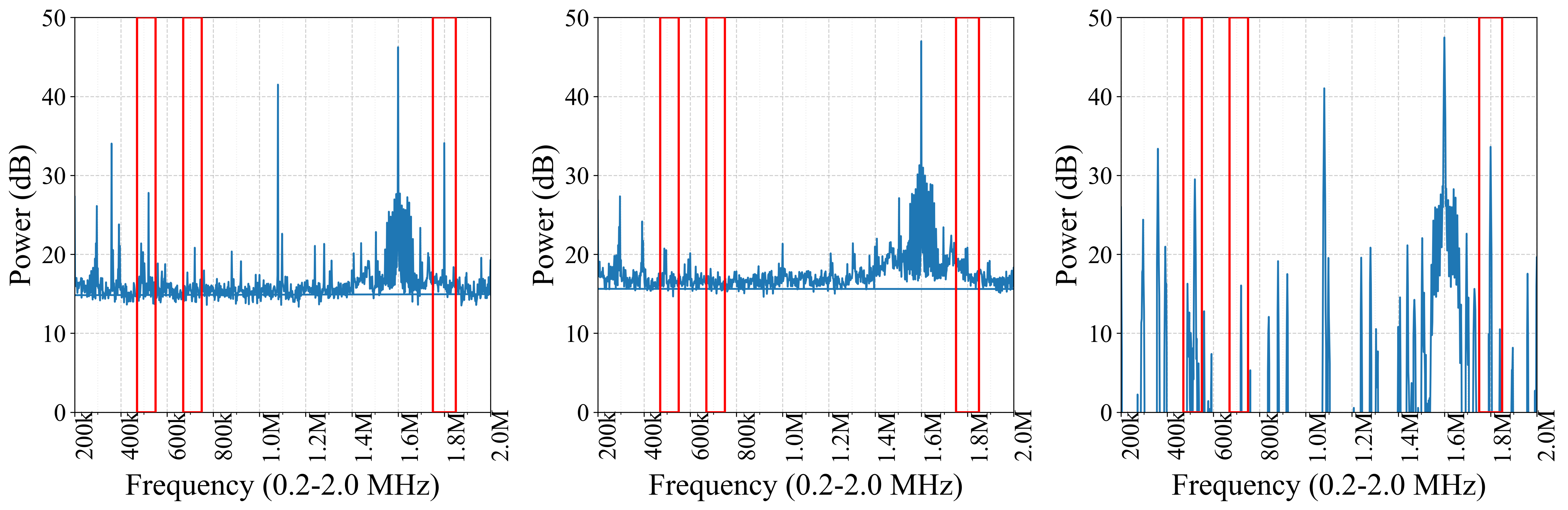} % 替换为实际文件名
    \caption{Visualization of original signal spectrum, ambient noise spectrum, and gesture spectrum after denoising.}
    \label{SignalVisualization}
\end{figure}

After generating the average power spectra of both gesture and ambient noise data, we applied the denoising method described in Section~\ref{sec:Noisereduction}. The results are shown in Fig.~\ref{SignalVisualization}. From left to right, the figure presents: (i) the spectrum containing gesture signals and noise, (ii) the spectrum of ambient noise, and (iii) the spectrum of a gesture signal after denoising. Prior to processing, the entire signal environment contained approximately 18\,dB of background noise, which blurred the distinction between gestures. After VMD-based decomposition and spectral subtraction, gesture-specific features became far more salient. Notably, different gestures induced distinct attenuation patterns across frequency bands, providing reliable features for classification. These enhanced features are subsequently used as input for the machine learning model.  

\textbf{Effect of Denoising on Accuracy.}  
We next quantified the contribution of our denoising method through an ablation study using the baseline hardware configuration with a Xiaomi 10 smartphone. Two experiments were conducted: (1) the full pipeline with VMD-based spectral subtraction, and (2) a baseline where raw, noisy spectra were directly used for classification. As shown in Table~\ref{tab:vmd-single}, the improvement of 8.43 percentage points confirms that denoising substantially enhances recognition accuracy by clarifying gesture features, making it a critical component of our pipeline.  

\begin{table}[tbp]
\centering
\caption{The impact of noise reduction on accuracy}
\label{tab:vmd-single}
\color{black}
\begin{tabular}{l r}
\toprule
\textbf{Model} & \textbf{Classification Accuracy} \\
\midrule
With VMD Denoising & 99.20\% \\
Without VMD Denoising & 90.77\% \\
\bottomrule
\end{tabular}
\end{table}

\subsection{Active Modulation via Negative Feedback}
\label{sec:feedback}
As a proof-of-concept for future work, we explored whether the negative feedback loop of a wireless charger could be exploited to actively modulate a stable, characteristic signal, rather than relying solely on passive gesture perturbations.

A typical wireless charger consists of a transmitter (Tx) and a receiver (Rx) that exchange energy through electromagnetic induction or magnetic resonance ~\cite{la2021wireless}. The transmitter must continuously regulate its power output: excessive power risks overheating, inefficiency, or battery damage, while insufficient power leads to slow charging. To maintain stability, the system monitors the receiver’s load through a feedback loop. Whenever the output deviates from the desired level, the transmitter automatically adjusts its driving signal (e.g., oscillation frequency, duty cycle, or voltage amplitude) to restore balance.

To test whether this feedback mechanism could be harnessed for active modulation, we developed a program that deliberately oscillates the phone’s power consumption. The CPU alternates between a “busy” compute state and an “idle” wait state at a designated frequency $f$. We hypothesized that this periodic load variation at the Rx would propagate back to the Tx via the feedback loop, thereby embedding a detectable modulation into the transmitted EM field.

\begin{figure}[tbp]
  \centering
  \includegraphics[width=0.5\linewidth]{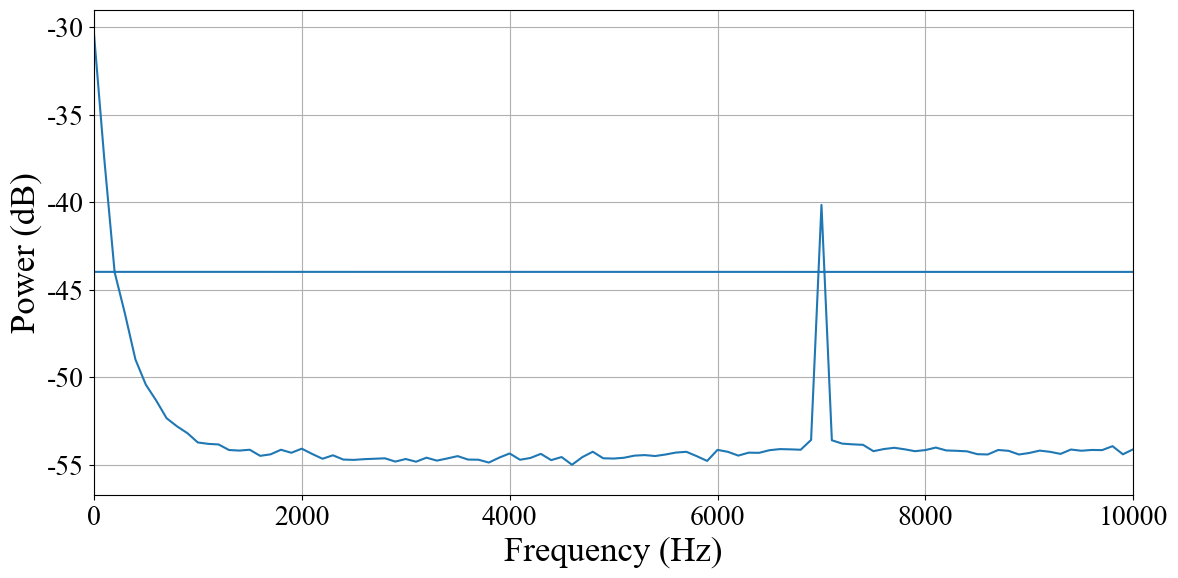} 
  \caption{Detection of a software-induced 7 kHz modulation amplified by the wireless charger’s feedback loop.}
  \label{fig:7kSignal}
\end{figure}

We conducted an experiment with the CPU load oscillating at 7 kHz. Using the sound card setup as a sensitive audio-band data acquisition equipment, we monitored the resulting EM field. As shown in Fig.~\ref{fig:7kSignal}, we successfully detected a strong, stable 7 kHz signal emitted by the charger, directly synchronized with the CPU load oscillation. This confirms the feasibility of intentionally modulating the charger’s EM field via software control. In future work, we plan to leverage this mechanism to generate actively controllable reference signals, opening the door to more reliable and robust gesture recognition.

%% file: userstudy.tex
\section{User Study}
To evaluate the user acceptance and satisfaction of EMGesture, we conducted a study with 20 participants (N=20; aged 18–45; 50\% male, 50\% female). The participants represented a diverse range of occupations, including students, office workers, and freelancers.

\textbf{Procedure.} Participants first received a short training session to familiarize themselves with the system and practiced the eight predefined gestures until they demonstrated proficiency. To support the study, we developed a companion app that provided real-time feedback by displaying the gesture recognized by the system. After completing the task (performing all eight gestures), participants filled out a five-point Likert-scale questionnaire and participated in a semi-structured interview. The questionnaire assessed five key metrics: \textit{Novelty}, \textit{Ease of Use (Usability)}, \textit{Learnability}, \textit{Naturalness of Interaction}, and \textit{Future Use Intention}.

\begin{figure}[tbp]
  \centering
  \includegraphics[width=0.7\linewidth]{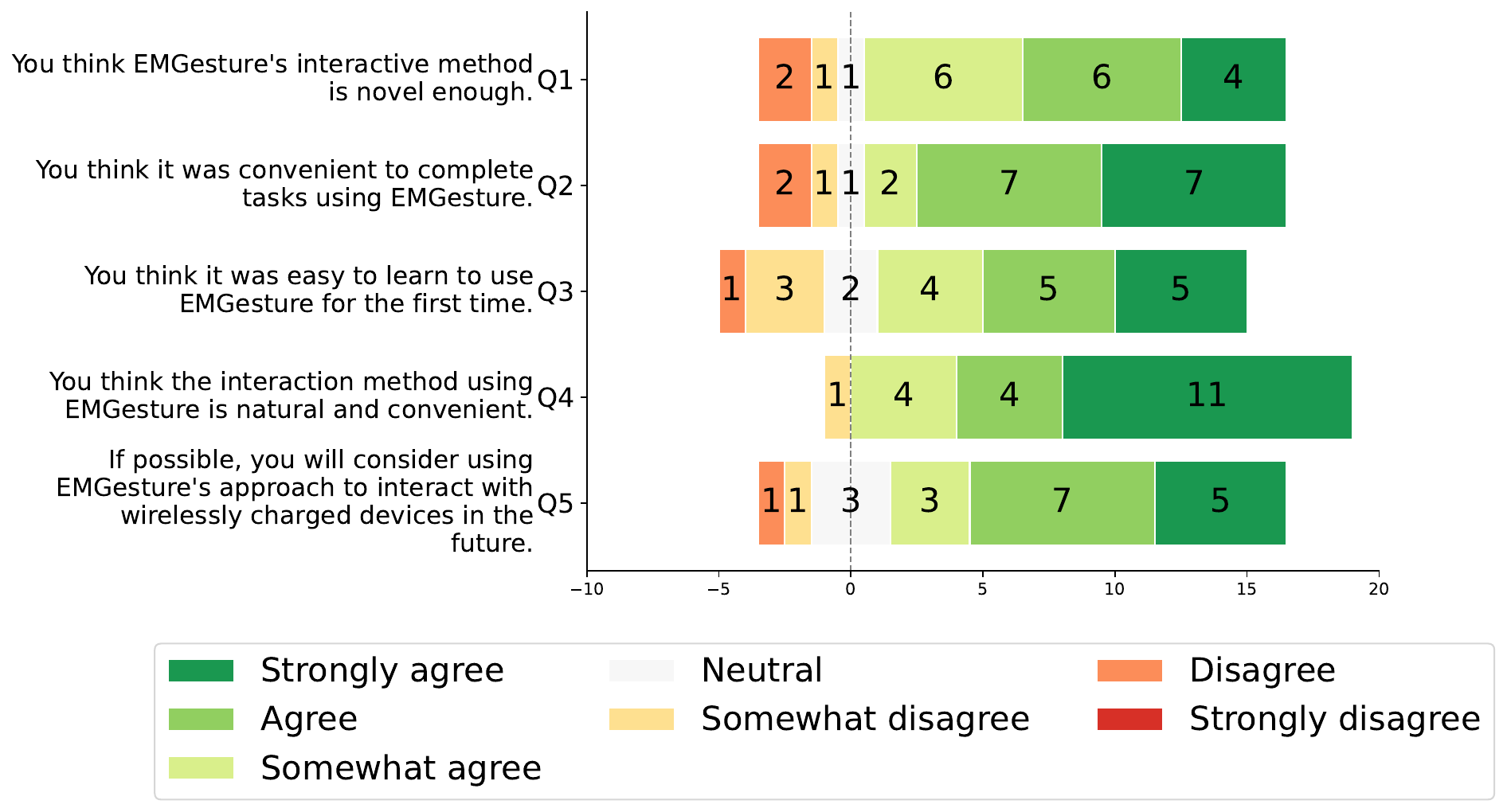} 
  \caption{Post-survey results on usability and acceptance of EMGesture}
  \label{fig:Survey}
\end{figure}

\textbf{Results.} The survey results are presented in Fig.~\ref{fig:Survey}.  
In terms of \textit{Novelty}, 80\% of participants rated EMGesture as novel to varying degrees. For \textit{Usability}, 80\% of participants found the system easy to operate, with one participant (P9) noting that it “required almost no additional instructions to understand.” Regarding \textit{Learnability}, most participants reported that the gestures were easy to learn, although 10\% found it initially challenging, likely due to their lack of prior experience. In terms of \textit{Naturalness of Interaction}, more than half of the participants rated the method as “very natural and smooth,” with P11 remarking that it was “similar to what they were already used to.” Finally, for \textit{Future Use Intention}, 75\% of participants expressed willingness to consider adopting EMGesture for smart home or in-car control scenarios.

\textbf{Qualitative Feedback.} In addition to the questionnaire, interviews provided further insights:  
\textit{P12 (28 years old):} “I find it very interesting, especially when I imagine controlling music in the car without looking at the screen. It’s very convenient.”  
\textit{P23 (34 years old):} “It feels more practical than camera recognition because it doesn’t raise privacy concerns and doesn’t rely on lighting.”  
\textit{P37 (22 years old):} “I thought it would be difficult to learn at first, but it only took a minute or two to get used to. It feels similar to everyday phone gestures.”  
\textit{P45 (41 years old):} “Maintaining gestures for a long time can be a bit tiring, but for everyday use the operations are short, so it’s acceptable.”

\textbf{Summary.} Overall, the user study demonstrates that \textit{EMGesture} performs well in terms of novelty, usability, and user acceptance. Participants considered the interaction method natural and intuitive, imposing minimal physical or cognitive effort. Supported by both quantitative and qualitative feedback, these findings indicate that \textit{EMGesture} has strong potential for real-world adoption, particularly in smart home and in-car control scenarios.

%% file: discussion.tex
\section{Discussion}
Our study demonstrates the feasibility of leveraging electromagnetic (EM) signals from wireless chargers for gesture-based interaction. While the evaluation confirms high accuracy, robustness across devices, and strong user acceptance, several considerations merit further discussion.

\textbf{System limitations.} First, our current prototype requires an external antenna, amplifier, and SDR platform. Although we have argued for future miniaturization and integration into wireless chargers, additional engineering efforts are needed to realize a compact, consumer-ready solution. Second, while the system performs robustly across different environments and devices, our experiments were limited to a relatively small set of charger and phone models. Broader evaluations in diverse real-world conditions (e.g., varying furniture layouts, denser EM environments) are needed to ensure generalizability.

\textbf{Gesture design and usability.} Our study focused on a predefined set of eight gestures. While these gestures were sufficient for validating feasibility, the range of supported interactions could be expanded. Designing gestures that are both intuitive and ergonomically comfortable for prolonged use remains an open challenge. Additionally, while most participants reported low physical and cognitive effort, some noted mild fatigue during sustained interactions, suggesting that gesture duration and complexity should be carefully optimized.

\textbf{Signal processing and modeling.} We adopted Variational Mode Decomposition (VMD) combined with spectral subtraction as our denoising pipeline, and Random Forests for classification. This choice yielded robust performance while remaining computationally lightweight. However, alternative denoising techniques (e.g., deep learning–based denoisers) or advanced classifiers (e.g., convolutional or transformer-based models) may further improve accuracy, particularly for more complex gesture sets. At the same time, exploring such methods must balance computational cost and deployability.

\textbf{Future opportunities.} Beyond passive gesture sensing, our proof-of-concept experiment demonstrated the feasibility of actively modulating charger signals via negative feedback. This opens promising opportunities for richer, controllable EM interactions. Moreover, as highlighted in prior work, wireless charging is not limited to smartphones but increasingly extends to wearables, electric vehicles, and smart appliances. Building on this broader applicability, our approach could be extended to other domains—such as public interfaces, automotive systems, or assistive technologies—further broadening the impact of EM-based interaction.

%% file: conclusion.tex
\section{Conclusion}
In this work, we presented \textit{EMGesture}, a novel interaction system that leverages the strong ambient electromagnetic signals emitted by wireless chargers to enable robust, contact-free gesture recognition. Through a carefully designed signal processing pipeline—including short-window FFT averaging, VMD-based denoising, and Random Forest classification—our method achieves high accuracy while remaining computationally lightweight. 

We implemented a prototype system and conducted extensive evaluations across diverse users, devices, chargers, and environments. The results consistently demonstrated strong recognition accuracy (up to 97.59\%), resilience to ambient electromagnetic interference, and user acceptance confirmed by a controlled user study. Furthermore, our proof-of-concept experiment on charger negative feedback highlights the potential for extending EM-based interaction from passive sensing to active modulation.

We provide a viable and pervasive interaction technique that eliminates the common limitations of conventional methods, such as contact requirements, privacy concerns, and high costs associated with camera-based or sensor-based solutions
, offering a practical, privacy-conscious, and cost-effective solution.

In the future, we plan to integrate the entire sensing pipeline into compact hardware for seamless deployment, and expanding application domains beyond smartphones to smart homes, automotive systems, and assistive technologies.

%% file: main.bbl

\begin{thebibliography}{50}

%%% the Web address.

\ifx \showCODEN    \undefined \def \showCODEN     #1{\unskip}     \fi
\ifx \showISBNx    \undefined \def \showISBNx     #1{\unskip}     \fi
\ifx \showISBNxiii \undefined \def \showISBNxiii  #1{\unskip}     \fi
\ifx \showISSN     \undefined \def \showISSN      #1{\unskip}     \fi
\ifx \showLCCN     \undefined \def \showLCCN      #1{\unskip}     \fi
\ifx \shownote     \undefined \def \shownote      #1{#1}          \fi
\ifx \showarticletitle \undefined \def \showarticletitle #1{#1}   \fi
\ifx \showURL      \undefined \def \showURL       {\relax}        \fi
% invisible to TeX
\providecommand\bibfield[2]{#2}
\providecommand\bibinfo[2]{#2}
\providecommand\natexlab[1]{#1}
\providecommand\showeprint[2][]{arXiv:#2}

\bibitem[PDF(2021)]%
        {PDFSmart27:online}
 \bibinfo{year}{2021}\natexlab{}.
\newblock \bibinfo{title}{(PDF) Smart Living: An Interactive Control System for
  Household Appliances}.
\newblock
\urldef\tempurl%
\url{https://www.researchgate.net/publication/348796023\_Smart\_Living\_An\_Interactive\_Control\_System\_for\_Household\_Appliances}
\showURL{%
\tempurl}


\bibitem[Dis(2023)]%
        {Distract53:online}
 \bibinfo{year}{2023}\natexlab{}.
\newblock \bibinfo{title}{Distracted Driving Dangers and Statistics | NHTSA}.
\newblock
\urldef\tempurl%
\url{https://www.nhtsa.gov/risky-driving/distracted-driving}
\showURL{%
\tempurl}


\bibitem[IoT(2023)]%
        {IoTSecur47:online}
 \bibinfo{year}{2023}\natexlab{}.
\newblock \bibinfo{title}{IoT Security Cameras are Vulnerable to Cyberattacks}.
\newblock
\urldef\tempurl%
\url{https://asimily.com/blog/iot-security-cameras-are-vulnerable-cyberattacks}
\showURL{%
\tempurl}


\bibitem[Hum(2025)]%
        {HumanCom90:online}
 \bibinfo{year}{2025}\natexlab{}.
\newblock \bibinfo{title}{Human Computer Interaction Market Report 2025 - Stats
  \& Trends}.
\newblock
\urldef\tempurl%
\url{https://www.thebusinessresearchcompany.com/report/human-computer-interaction-global-market-report}
\showURL{%
\tempurl}
\newblock
\shownote{[Online; accessed 2025-09-10]}.


\bibitem[The(2025)]%
        {TheVoice1:online}
 \bibinfo{year}{2025}\natexlab{}.
\newblock \bibinfo{title}{The Voice Privacy Problem}.
\newblock
\urldef\tempurl%
\url{https://www.kardome.com/blog-posts/voice-privacy-concerns}
\showURL{%
\tempurl}


\bibitem[Wha(2025)]%
        {Whatares45:online}
 \bibinfo{year}{2025}\natexlab{}.
\newblock \bibinfo{title}{What are some possible causes of interference on PCAP
  screens? \- Elo \- Technical Support}.
\newblock
\urldef\tempurl%
\url{https://elosupport.elotouch.com/hc/en-us/articles/31651505678615-What-are-some-possible-causes-of-interference-on-PCAP-screens}
\showURL{%
\tempurl}


\bibitem[Abdelnasser et~al\mbox{.}(2020)]%
        {abdelnasser2020magstroke}
\bibfield{author}{\bibinfo{person}{Heba Abdelnasser}, \bibinfo{person}{Khaled~A
  Harras}, {and} \bibinfo{person}{Moustafa Youssef}.}
  \bibinfo{year}{2020}\natexlab{}.
\newblock \showarticletitle{Magstroke: A magnetic based virtual keyboard for
  off-the-shelf smart devices}. In \bibinfo{booktitle}{\emph{2020 17th Annual
  IEEE International Conference on Sensing, Communication, and Networking
  (SECON)}}. IEEE, \bibinfo{pages}{1\--\--9}.
\newblock


\bibitem[Balanis(2012)]%
        {balanis2012advanced}
\bibfield{author}{\bibinfo{person}{Constantine~A Balanis}.}
  \bibinfo{year}{2012}\natexlab{}.
\newblock \bibinfo{booktitle}{\emph{Advanced engineering electromagnetics}}.
\newblock \bibinfo{publisher}{John Wiley \& Sons}.
\newblock


\bibitem[Breiman(2001)]%
        {breiman2001random}
\bibfield{author}{\bibinfo{person}{Leo Breiman}.}
  \bibinfo{year}{2001}\natexlab{}.
\newblock \showarticletitle{Random forests}.
\newblock \bibinfo{journal}{\emph{Machine learning}} \bibinfo{volume}{45},
  \bibinfo{number}{1} (\bibinfo{year}{2001}), \bibinfo{pages}{5\--\--32}.
\newblock


\bibitem[Chen et~al\mbox{.}(2022)]%
        {chen2022emd}
\bibfield{author}{\bibinfo{person}{Tao Chen}, \bibinfo{person}{Shuncheng Gao},
  \bibinfo{person}{Shilian Zheng}, \bibinfo{person}{Shanqing Yu},
  \bibinfo{person}{Qi Xuan}, \bibinfo{person}{Caiyi Lou}, {and}
  \bibinfo{person}{Xiaoniu Yang}.} \bibinfo{year}{2022}\natexlab{}.
\newblock \showarticletitle{EMD and VMD empowered deep learning for radio
  modulation recognition}.
\newblock \bibinfo{journal}{\emph{IEEE Transactions on Cognitive Communications
  and Networking}} \bibinfo{volume}{9}, \bibinfo{number}{1}
  (\bibinfo{year}{2022}), \bibinfo{pages}{43--57}.
\newblock


\bibitem[Dai et~al\mbox{.}(2025)]%
        {dai2025robust}
\bibfield{author}{\bibinfo{person}{Zhenlei Dai}, \bibinfo{person}{Liangchen
  Hu}, {and} \bibinfo{person}{Huaijiang Sun}.} \bibinfo{year}{2025}\natexlab{}.
\newblock \showarticletitle{Robust generalized PCA for enhancing
  discriminability and recoverability}.
\newblock \bibinfo{journal}{\emph{Neural Networks}}  \bibinfo{volume}{181}
  (\bibinfo{year}{2025}), \bibinfo{pages}{106814}.
\newblock


\bibitem[Dragomiretskiy and Zosso(2013)]%
        {dragomiretskiy2013variational}
\bibfield{author}{\bibinfo{person}{Konstantin Dragomiretskiy} {and}
  \bibinfo{person}{Dominique Zosso}.} \bibinfo{year}{2013}\natexlab{}.
\newblock \showarticletitle{Variational mode decomposition}.
\newblock \bibinfo{journal}{\emph{IEEE transactions on signal processing}}
  \bibinfo{volume}{62}, \bibinfo{number}{3} (\bibinfo{year}{2013}),
  \bibinfo{pages}{531\--\--544}.
\newblock


\bibitem[Gao et~al\mbox{.}(2021)]%
        {gao2021towards}
\bibfield{author}{\bibinfo{person}{Ruiyang Gao}, \bibinfo{person}{Mi Zhang},
  \bibinfo{person}{Jie Zhang}, \bibinfo{person}{Yang Li}, \bibinfo{person}{Enze
  Yi}, \bibinfo{person}{Dan Wu}, \bibinfo{person}{Leye Wang}, {and}
  \bibinfo{person}{Daqing Zhang}.} \bibinfo{year}{2021}\natexlab{}.
\newblock \showarticletitle{Towards position\-independent sensing for gesture
  recognition with Wi-Fi}.
\newblock \bibinfo{journal}{\emph{Proceedings of the ACM on Interactive,
  Mobile, Wearable and Ubiquitous Technologies}} \bibinfo{volume}{5},
  \bibinfo{number}{2} (\bibinfo{year}{2021}), \bibinfo{pages}{1\--\--28}.
\newblock


\bibitem[Gao et~al\mbox{.}(2024)]%
        {gao2024parameter}
\bibfield{author}{\bibinfo{person}{Ziqi Gao}, \bibinfo{person}{Qichao Wang},
  \bibinfo{person}{Aochuan Chen}, \bibinfo{person}{Zijing Liu},
  \bibinfo{person}{Bingzhe Wu}, \bibinfo{person}{Liang Chen}, {and}
  \bibinfo{person}{Jia Li}.} \bibinfo{year}{2024}\natexlab{}.
\newblock \showarticletitle{Parameter-efficient fine-tuning with discrete
  fourier transform}.
\newblock \bibinfo{journal}{\emph{arXiv preprint arXiv:2405.03003}}
  (\bibinfo{year}{2024}).
\newblock


\bibitem[Hayashi et~al\mbox{.}(2021)]%
        {hayashi2021radarnet}
\bibfield{author}{\bibinfo{person}{Eiji Hayashi}, \bibinfo{person}{Jaime Lien},
  \bibinfo{person}{Nicholas Gillian}, \bibinfo{person}{Leonardo Giusti},
  \bibinfo{person}{Dave Weber}, \bibinfo{person}{Jin Yamanaka},
  \bibinfo{person}{Lauren Bedal}, {and} \bibinfo{person}{Ivan Poupyrev}.}
  \bibinfo{year}{2021}\natexlab{}.
\newblock \showarticletitle{Radarnet: Efficient gesture recognition technique
  utilizing a miniature radar sensor}. In \bibinfo{booktitle}{\emph{Proceedings
  of the 2021 CHI Conference on Human Factors in Computing Systems}}.
  \bibinfo{pages}{1\--\--14}.
\newblock


\bibitem[Hu and Szymczak(2023)]%
        {hu2023review}
\bibfield{author}{\bibinfo{person}{Jianchang Hu} {and} \bibinfo{person}{Silke
  Szymczak}.} \bibinfo{year}{2023}\natexlab{}.
\newblock \showarticletitle{A review on longitudinal data analysis with random
  forest}.
\newblock \bibinfo{journal}{\emph{Briefings in bioinformatics}}
  \bibinfo{volume}{24}, \bibinfo{number}{2} (\bibinfo{year}{2023}),
  \bibinfo{pages}{bbad002}.
\newblock


\bibitem[IN(2025)]%
        {HowtoChe84:online}
\bibfield{author}{\bibinfo{person}{Dell IN}.} \bibinfo{year}{2025}\natexlab{}.
\newblock \bibinfo{title}{How to Check for WiFi Interference and Reduce
  Wireless Signal Problems - Dell India}.
\newblock
\urldef\tempurl%
\url{https://www.dell.com/support/kbdoc/en-in/000150359/how-to-identify-and-reduce-wireless-signal-interference}
\showURL{%
\tempurl}


\bibitem[infilityfun@gmail.com(2025)]%
        {65Intera48:online}
\bibfield{author}{\bibinfo{person}{infilityfun@gmail.com}.}
  \bibinfo{year}{2025}\natexlab{}.
\newblock \bibinfo{title}{Interactive Touch Screen Price in 2025 | Cost \&
  Buying Guide}.
\newblock
\urldef\tempurl%
\url{https://touchwo.com/65-inch-interactive-touch-screen-price}
\showURL{%
\tempurl}


\bibitem[Jin et~al\mbox{.}(2024)]%
        {jin2024rodar}
\bibfield{author}{\bibinfo{person}{Can Jin}, \bibinfo{person}{Xiangzhu Meng},
  \bibinfo{person}{Xuanheng Li}, \bibinfo{person}{Jie Wang},
  \bibinfo{person}{Miao Pan}, {and} \bibinfo{person}{Yuguang Fang}.}
  \bibinfo{year}{2024}\natexlab{}.
\newblock \showarticletitle{Rodar: Robust gesture recognition based on mmWave
  radar under human activity interference}.
\newblock \bibinfo{journal}{\emph{IEEE Transactions on Mobile Computing}}
  \bibinfo{volume}{23}, \bibinfo{number}{12} (\bibinfo{year}{2024}),
  \bibinfo{pages}{11735\--\--11749}.
\newblock


\bibitem[Khanna et~al\mbox{.}(2024)]%
        {khanna2024hand}
\bibfield{author}{\bibinfo{person}{Prerna Khanna}, \bibinfo{person}{IV
  Ramakrishnan}, \bibinfo{person}{Shubham Jain}, \bibinfo{person}{Xiaojun Bi},
  {and} \bibinfo{person}{Aruna Balasubramanian}.}
  \bibinfo{year}{2024}\natexlab{}.
\newblock \showarticletitle{Hand gesture recognition for blind users by
  tracking 3d gesture trajectory}. In \bibinfo{booktitle}{\emph{Proceedings of
  the 2024 CHI Conference on Human Factors in Computing Systems}}.
  \bibinfo{pages}{1\--\--15}.
\newblock


\bibitem[La~Cour et~al\mbox{.}(2021)]%
        {la2021wireless}
\bibfield{author}{\bibinfo{person}{Alexander~S La~Cour},
  \bibinfo{person}{Khurram~K Afridi}, {and} \bibinfo{person}{G~Edward Suh}.}
  \bibinfo{year}{2021}\natexlab{}.
\newblock \showarticletitle{Wireless charging power side\-channel attacks}. In
  \bibinfo{booktitle}{\emph{Proceedings of the 2021 ACM SIGSAC Conference on
  Computer and Communications Security}}. \bibinfo{pages}{651\--\--665}.
\newblock


\bibitem[Li et~al\mbox{.}(2022)]%
        {li2022room}
\bibfield{author}{\bibinfo{person}{Dong Li}, \bibinfo{person}{Jialin Liu},
  \bibinfo{person}{Sunghoon~Ivan Lee}, {and} \bibinfo{person}{Jie Xiong}.}
  \bibinfo{year}{2022}\natexlab{}.
\newblock \showarticletitle{Room\-scale hand gesture recognition using smart
  speakers}. In \bibinfo{booktitle}{\emph{Proceedings of the 20th ACM
  Conference on Embedded Networked Sensor Systems}}.
  \bibinfo{pages}{462\--\--475}.
\newblock


\bibitem[Li et~al\mbox{.}(2024)]%
        {li2024magnetic}
\bibfield{author}{\bibinfo{person}{Jiachun Li}, \bibinfo{person}{Yan Meng},
  \bibinfo{person}{Le Zhang}, \bibinfo{person}{Guoxing Chen},
  \bibinfo{person}{Yuan Tian}, {and} \bibinfo{person}{Haojin Zhu}.}
  \bibinfo{year}{2024}\natexlab{}.
\newblock \showarticletitle{A Magnetic Signal Based Device Fingerprinting
  Scheme in Wireless Charging}.
\newblock \bibinfo{journal}{\emph{IEEE Transactions on Dependable and Secure
  Computing}} (\bibinfo{year}{2024}).
\newblock


\bibitem[Li et~al\mbox{.}(2023)]%
        {li2023audiosense}
\bibfield{author}{\bibinfo{person}{Yijie Li}, \bibinfo{person}{Xiatong Tong},
  \bibinfo{person}{Qianfei Ren}, \bibinfo{person}{Qingyang Li},
  \bibinfo{person}{Lanqing Yang}, \bibinfo{person}{Yi-Chao Chen},
  \bibinfo{person}{Guangtao Xue}, \bibinfo{person}{Xiaoyu Ji}, {and}
  \bibinfo{person}{Jiadi Yu}.} \bibinfo{year}{2023}\natexlab{}.
\newblock \showarticletitle{AUDIOSENSE: Leveraging Current to Acoustic Channel
  to Detect Appliances at Single-Point}. In \bibinfo{booktitle}{\emph{2023 20th
  Annual IEEE International Conference on Sensing, Communication, and
  Networking (SECON)}}. IEEE, \bibinfo{pages}{240\--\--248}.
\newblock


\bibitem[Liu et~al\mbox{.}(2024)]%
        {liu2024unifi}
\bibfield{author}{\bibinfo{person}{Yan Liu}, \bibinfo{person}{Anlan Yu},
  \bibinfo{person}{Leye Wang}, \bibinfo{person}{Bin Guo}, \bibinfo{person}{Yang
  Li}, \bibinfo{person}{Enze Yi}, {and} \bibinfo{person}{Daqing Zhang}.}
  \bibinfo{year}{2024}\natexlab{}.
\newblock \showarticletitle{Unifi: A unified framework for generalizable
  gesture recognition with wi-fi signals using consistency\-guided multi\-view
  networks}.
\newblock \bibinfo{journal}{\emph{Proceedings of the ACM on Interactive,
  Mobile, Wearable and Ubiquitous Technologies}} \bibinfo{volume}{7},
  \bibinfo{number}{4} (\bibinfo{year}{2024}), \bibinfo{pages}{1\--\--29}.
\newblock


\bibitem[Liu et~al\mbox{.}(2023)]%
        {liu2023camradar}
\bibfield{author}{\bibinfo{person}{Ziwei Liu}, \bibinfo{person}{Feng Lin},
  \bibinfo{person}{Chao Wang}, \bibinfo{person}{Yijie Shen},
  \bibinfo{person}{Zhongjie Ba}, \bibinfo{person}{Li Lu},
  \bibinfo{person}{Wenyao Xu}, {and} \bibinfo{person}{Kui Ren}.}
  \bibinfo{year}{2023}\natexlab{}.
\newblock \showarticletitle{Camradar: Hidden camera detection leveraging
  amplitude-modulated sensor images embedded in electromagnetic emanations}.
\newblock \bibinfo{journal}{\emph{Proceedings of the ACM on Interactive,
  Mobile, Wearable and Ubiquitous Technologies}} \bibinfo{volume}{6},
  \bibinfo{number}{4} (\bibinfo{year}{2023}), \bibinfo{pages}{1\--\--25}.
\newblock


\bibitem[Ma et~al\mbox{.}(2023)]%
        {ma2023mimo}
\bibfield{author}{\bibinfo{person}{Wenyan Ma}, \bibinfo{person}{Lipeng Zhu},
  {and} \bibinfo{person}{Rui Zhang}.} \bibinfo{year}{2023}\natexlab{}.
\newblock \showarticletitle{MIMO capacity characterization for movable antenna
  systems}.
\newblock \bibinfo{journal}{\emph{IEEE Transactions on Wireless
  Communications}} \bibinfo{volume}{23}, \bibinfo{number}{4}
  (\bibinfo{year}{2023}), \bibinfo{pages}{3392--3407}.
\newblock


\bibitem[Mohamed et~al\mbox{.}(2024)]%
        {mohamed2024wireless}
\bibfield{author}{\bibinfo{person}{Ahmed~AS Mohamed}, \bibinfo{person}{Ahmed~A
  Shaier}, \bibinfo{person}{Hamid Metwally}, {and} \bibinfo{person}{Sameh~I
  Selem}.} \bibinfo{year}{2024}\natexlab{}.
\newblock \showarticletitle{Wireless charging technologies for electric
  vehicles: Inductive, capacitive, and magnetic gear}.
\newblock \bibinfo{journal}{\emph{IET Power Electronics}} \bibinfo{volume}{17},
  \bibinfo{number}{16} (\bibinfo{year}{2024}), \bibinfo{pages}{3139\--\--3165}.
\newblock


\bibitem[Mohammadzadeh et~al\mbox{.}(2022)]%
        {mohammadzadeh2022robust}
\bibfield{author}{\bibinfo{person}{Saeed Mohammadzadeh},
  \bibinfo{person}{V{\'\i}tor~H Nascimento}, \bibinfo{person}{Rodrigo~C
  De~Lamare}, {and} \bibinfo{person}{Osman Kukrer}.}
  \bibinfo{year}{2022}\natexlab{}.
\newblock \showarticletitle{Robust adaptive beamforming based on power method
  processing and spatial spectrum matching}. In
  \bibinfo{booktitle}{\emph{ICASSP 2022-2022 IEEE International Conference on
  Acoustics, Speech and Signal Processing (ICASSP)}}. IEEE,
  \bibinfo{pages}{4903--4907}.
\newblock


\bibitem[Ni et~al\mbox{.}(2023)]%
        {ni2023uncovering}
\bibfield{author}{\bibinfo{person}{Tao Ni}, \bibinfo{person}{Xiaokuan Zhang},
  \bibinfo{person}{Chaoshun Zuo}, \bibinfo{person}{Jianfeng Li},
  \bibinfo{person}{Zhenyu Yan}, \bibinfo{person}{Wubing Wang},
  \bibinfo{person}{Weitao Xu}, \bibinfo{person}{Xiapu Luo}, {and}
  \bibinfo{person}{Qingchuan Zhao}.} \bibinfo{year}{2023}\natexlab{}.
\newblock \showarticletitle{Uncovering user interactions on smartphones via
  contactless wireless charging side channels}. In
  \bibinfo{booktitle}{\emph{2023 IEEE Symposium on Security and Privacy (SP)}}.
  IEEE, \bibinfo{pages}{3399\--\--3415}.
\newblock


\bibitem[Online(2025)]%
        {Smarthom87:online}
\bibfield{author}{\bibinfo{person}{People's~Daily Online}.}
  \bibinfo{year}{2025}\natexlab{}.
\newblock \bibinfo{title}{Smart home appliances become new darlings of young
  consumers People's Daily Online}.
\newblock
\urldef\tempurl%
\url{https://en.people.cn/n3/2025/0324/c98649-20292974.html}
\showURL{%
\tempurl}


\bibitem[Palam{\`a} et~al\mbox{.}(2024)]%
        {palama20245g}
\bibfield{author}{\bibinfo{person}{Ivan Palam{\`a}}, \bibinfo{person}{Yago
  Lizarribar}, \bibinfo{person}{Lorenzo~Maria Monteforte},
  \bibinfo{person}{Giuseppe Santaromita}, \bibinfo{person}{Stefania
  Bartoletti}, \bibinfo{person}{Domenico Giustiniano},
  \bibinfo{person}{Giuseppe Bianchi}, {and} \bibinfo{person}{Nicola~Blefari
  Melazzi}.} \bibinfo{year}{2024}\natexlab{}.
\newblock \showarticletitle{5G positioning with software-defined radios}.
\newblock \bibinfo{journal}{\emph{Computer Networks}}  \bibinfo{volume}{250}
  (\bibinfo{year}{2024}), \bibinfo{pages}{110595}.
\newblock


\bibitem[Poojithainakota(2023)]%
        {ThePower76:online}
\bibfield{author}{\bibinfo{person}{Poojithainakota}.}
  \bibinfo{year}{2023}\natexlab{}.
\newblock \bibinfo{title}{The Power of Buttons: Beginners Guide for Seamless
  User Interaction | by Poojithainakota | Medium}.
\newblock
\urldef\tempurl%
\url{https://medium.com/@poojithainakota/the-power-of-buttons-beginners-guide-for-seamless-user-interaction-e0efb60fee64}
\showURL{%
\tempurl}


\bibitem[Por et~al\mbox{.}(2019)]%
        {por2019nyquist}
\bibfield{author}{\bibinfo{person}{Emiel Por}, \bibinfo{person}{Maaike
  Van~Kooten}, {and} \bibinfo{person}{Vanja Sarkovic}.}
  \bibinfo{year}{2019}\natexlab{}.
\newblock \showarticletitle{Nyquist\-\-Shannon sampling theorem}.
\newblock \bibinfo{journal}{\emph{Leiden University}} \bibinfo{volume}{1},
  \bibinfo{number}{1} (\bibinfo{year}{2019}), \bibinfo{pages}{1\--\--2}.
\newblock


\bibitem[Quora(2025)]%
        {1Whydoes87:online}
\bibfield{author}{\bibinfo{person}{Quora}.} \bibinfo{year}{2025}\natexlab{}.
\newblock \bibinfo{title}{Why does a touchscreen not work when it is wet or the
  fingers of user are wet? \- Quora}.
\newblock
\urldef\tempurl%
\url{https://www.quora.com/Why-does-a-touchscreen-not-work-when-it-is-wet-or-the-fingers-of-user-are-wet}
\showURL{%
\tempurl}


\bibitem[Rilling et~al\mbox{.}(2003)]%
        {rilling2003empirical}
\bibfield{author}{\bibinfo{person}{Gabriel Rilling}, \bibinfo{person}{Patrick
  Flandrin}, {and} \bibinfo{person}{Paulo Goncalves}.}
  \bibinfo{year}{2003}\natexlab{}.
\newblock \showarticletitle{On empirical mode decomposition and its
  algorithms}. In \bibinfo{booktitle}{\emph{IEEE\-EURASIP workshop on nonlinear
  signal and image processing NSIP-03}}.
\newblock


\bibitem[SanMiguel and Cavallaro(2016)]%
        {sanmiguel2016energy}
\bibfield{author}{\bibinfo{person}{Juan~Carlos SanMiguel} {and}
  \bibinfo{person}{Andrea Cavallaro}.} \bibinfo{year}{2016}\natexlab{}.
\newblock \showarticletitle{Energy consumption models for smart camera
  networks}.
\newblock \bibinfo{journal}{\emph{IEEE Transactions on Circuits and Systems for
  Video Technology}} \bibinfo{volume}{27}, \bibinfo{number}{12}
  (\bibinfo{year}{2016}), \bibinfo{pages}{2661\--\--2674}.
\newblock


\bibitem[Smith(2021)]%
        {smith2021spectral}
\bibfield{author}{\bibinfo{person}{Brian Smith}.}
  \bibinfo{year}{2021}\natexlab{}.
\newblock \showarticletitle{Spectral subtraction}.
\newblock  (\bibinfo{year}{2021}).
\newblock


\bibitem[Statista(2025)]%
        {Globalhu52:online}
\bibfield{author}{\bibinfo{person}{Statista}.} \bibinfo{year}{2025}\natexlab{}.
\newblock \bibinfo{title}{Global human machine interface market size 2026|
  Statista}.
\newblock
\urldef\tempurl%
\url{https://www.statista.com/statistics/1120222/global-human-machine-interface-market-size}
\showURL{%
\tempurl}


\bibitem[Swarztrauber(1984)]%
        {swarztrauber1984fft}
\bibfield{author}{\bibinfo{person}{Paul~N Swarztrauber}.}
  \bibinfo{year}{1984}\natexlab{}.
\newblock \showarticletitle{FFT algorithms for vector computers}.
\newblock \bibinfo{journal}{\emph{Parallel Comput.}} \bibinfo{volume}{1},
  \bibinfo{number}{1} (\bibinfo{year}{1984}), \bibinfo{pages}{45\--\--63}.
\newblock


\bibitem[Valkenborg et~al\mbox{.}(2023)]%
        {valkenborg2023support}
\bibfield{author}{\bibinfo{person}{Dirk Valkenborg}, \bibinfo{person}{Axel-Jan
  Rousseau}, \bibinfo{person}{Melvin Geubbelmans}, {and}
  \bibinfo{person}{Tomasz Burzykowski}.} \bibinfo{year}{2023}\natexlab{}.
\newblock \showarticletitle{Support vector machines}.
\newblock \bibinfo{journal}{\emph{American journal of orthodontics and
  dentofacial orthopedics}} \bibinfo{volume}{164}, \bibinfo{number}{5}
  (\bibinfo{year}{2023}), \bibinfo{pages}{754--757}.
\newblock


\bibitem[Van~Wageningen and Staring(2010)]%
        {van2010qi}
\bibfield{author}{\bibinfo{person}{Dries Van~Wageningen} {and}
  \bibinfo{person}{Toine Staring}.} \bibinfo{year}{2010}\natexlab{}.
\newblock \showarticletitle{The Qi wireless power standard}. In
  \bibinfo{booktitle}{\emph{Proceedings of 14th International Power Electronics
  and Motion Control Conference EPE-PEMC 2010}}. IEEE,
  \bibinfo{pages}{S15\--\--25}.
\newblock


\bibitem[Xu et~al\mbox{.}(2024)]%
        {xu2024gestureprint}
\bibfield{author}{\bibinfo{person}{Lilin Xu}, \bibinfo{person}{Keyi Wang},
  \bibinfo{person}{Chaojie Gu}, \bibinfo{person}{Xiuzhen Guo},
  \bibinfo{person}{Shibo He}, {and} \bibinfo{person}{Jiming Chen}.}
  \bibinfo{year}{2024}\natexlab{}.
\newblock \showarticletitle{GesturePrint: Enabling user identification for
  mmWave-based gesture recognition systems}. In \bibinfo{booktitle}{\emph{2024
  IEEE 44th International Conference on Distributed Computing Systems
  (ICDCS)}}. IEEE, \bibinfo{pages}{1074\--\--1085}.
\newblock


\bibitem[Xue et~al\mbox{.}(2021)]%
        {xue2021magnecomm+}
\bibfield{author}{\bibinfo{person}{Guangtao Xue}, \bibinfo{person}{Hao Pan},
  \bibinfo{person}{Yi-Chao Chen}, \bibinfo{person}{Xiaoyu Ji}, {and}
  \bibinfo{person}{Jiadi Yu}.} \bibinfo{year}{2021}\natexlab{}.
\newblock \showarticletitle{MagneComm+: Near\-field electromagnetic induction
  communication with magnetometer}.
\newblock \bibinfo{journal}{\emph{IEEE Transactions on Mobile Computing}}
  \bibinfo{volume}{22}, \bibinfo{number}{5} (\bibinfo{year}{2021}),
  \bibinfo{pages}{2789\--\--2801}.
\newblock


\bibitem[Yang et~al\mbox{.}(2024a)]%
        {yang2024maf}
\bibfield{author}{\bibinfo{person}{Yongjie Yang}, \bibinfo{person}{Tao Chen},
  \bibinfo{person}{Yujing Huang}, \bibinfo{person}{Xiuzhen Guo}, {and}
  \bibinfo{person}{Longfei Shangguan}.} \bibinfo{year}{2024}\natexlab{a}.
\newblock \showarticletitle{MAF: Exploring mobile acoustic field for
  hand-to-face gesture interactions}. In \bibinfo{booktitle}{\emph{Proceedings
  of the 2024 CHI Conference on Human Factors in Computing Systems}}.
  \bibinfo{pages}{1\--\--20}.
\newblock


\bibitem[Yang et~al\mbox{.}(2024b)]%
        {yang2024privacy}
\bibfield{author}{\bibinfo{person}{Yirui Yang}, \bibinfo{person}{Zihao Zhan},
  \bibinfo{person}{Honggang Yu}, \bibinfo{person}{Qinghui Huang}, {and}
  \bibinfo{person}{Shuo Wang}.} \bibinfo{year}{2024}\natexlab{b}.
\newblock \showarticletitle{A Privacy Leakage Issue in Qi\-compatible Cellphone
  Wireless Charging by Stray Magnetic Field Sniffing}. In
  \bibinfo{booktitle}{\emph{2024 IEEE Applied Power Electronics Conference and
  Exposition (APEC)}}. IEEE, \bibinfo{pages}{2870\--\--2876}.
\newblock


\bibitem[Zhan et~al\mbox{.}(2024)]%
        {zhan2024voltschemer}
\bibfield{author}{\bibinfo{person}{Zihao Zhan}, \bibinfo{person}{Yirui Yang},
  \bibinfo{person}{Haoqi Shan}, \bibinfo{person}{Hanqiu Wang},
  \bibinfo{person}{Yier Jin}, {and} \bibinfo{person}{Shuo Wang}.}
  \bibinfo{year}{2024}\natexlab{}.
\newblock \showarticletitle{\textit{VoltSchemer}: Use Voltage Noise to
  Manipulate Your Wireless Charger}. In \bibinfo{booktitle}{\emph{33rd USENIX
  Security Symposium (USENIX Security 24)}}. \bibinfo{pages}{3979\--\--3995}.
\newblock


\bibitem[Zhang et~al\mbox{.}(2024)]%
        {zhang2024adaptive}
\bibfield{author}{\bibinfo{person}{Junning Zhang}, \bibinfo{person}{Yicen Liu},
  \bibinfo{person}{Guoru Ding}, \bibinfo{person}{Bo Tang}, {and}
  \bibinfo{person}{Yanlong Chen}.} \bibinfo{year}{2024}\natexlab{}.
\newblock \showarticletitle{Adaptive decomposition and extraction network of
  individual fingerprint features for specific emitter identification}.
\newblock \bibinfo{journal}{\emph{IEEE Transactions on Information Forensics
  and Security}} (\bibinfo{year}{2024}).
\newblock


\bibitem[Zhang et~al\mbox{.}(2022)]%
        {zhang2022wi}
\bibfield{author}{\bibinfo{person}{Ronghui Zhang}, \bibinfo{person}{Chunxiao
  Jiang}, \bibinfo{person}{Sheng Wu}, \bibinfo{person}{Quan Zhou},
  \bibinfo{person}{Xiaojun Jing}, {and} \bibinfo{person}{Junsheng Mu}.}
  \bibinfo{year}{2022}\natexlab{}.
\newblock \showarticletitle{Wi\-Fi sensing for joint gesture recognition and
  human identification from few samples in human-computer interaction}.
\newblock \bibinfo{journal}{\emph{IEEE Journal on Selected Areas in
  Communications}} \bibinfo{volume}{40}, \bibinfo{number}{7}
  (\bibinfo{year}{2022}), \bibinfo{pages}{2193\--\--2205}.
\newblock


\bibitem[Zhang(2021)]%
        {zhang2021challenges}
\bibfield{author}{\bibinfo{person}{Shichao Zhang}.}
  \bibinfo{year}{2021}\natexlab{}.
\newblock \showarticletitle{Challenges in KNN classification}.
\newblock \bibinfo{journal}{\emph{IEEE Transactions on Knowledge and Data
  Engineering}} \bibinfo{volume}{34}, \bibinfo{number}{10}
  (\bibinfo{year}{2021}), \bibinfo{pages}{4663--4675}.
\newblock


\end{thebibliography}
